\documentclass[12pt]{article}
\usepackage{amsmath,amssymb}
\usepackage{epsfig}
\usepackage{graphicx}
\usepackage{a4}
\usepackage{latexsym}
\usepackage{cite}

\usepackage{color}
\usepackage{colordvi}
\usepackage{bm}
\DeclareGraphicsExtensions{.eps,.ps}

\usepackage{pslatex}
\usepackage{txfonts}
\usepackage[latin1]{inputenc}
\usepackage[T1]{fontenc}

\newcommand{\gsim}{\raisebox{-0.07cm}{$\:\:\stackrel{>}{{\scriptstyle \sim}}\:\: $} }
\newcommand{\lsim}{\raisebox{-0.07cm}{$\:\:\stackrel{<}{{\scriptstyle \sim}}\:\: $} }

\begin{document}

\setlength{\parskip}{0.2cm}
\setlength{\baselineskip}{0.535cm}

\begin{titlepage}
\noindent
DESY 09-102 \hfill {\tt arXiv:0908.2766 [hep-ph]}\\
SFB/CPP-09-072 \\
August 2009 
\hyphenation{neu-tri-no-nucleon}
\begin{center}
\Large {\bf The 3-, 4-, and 5-flavor NNLO Parton Distributions Functions}
\vspace*{2mm}
\Large{\bf from Deep-Inelastic-Scattering Data and at Hadron Colliders}
\\
\vspace{1cm}
\large
S. Alekhin$^{\, a,b}$\footnote{{\bf e-mail}: sergey.alekhin@ihep.ru},
J. Bl\"umlein $^{\, b,}$\footnote{{\bf e-mail}: johannes.bluemlein@desy.de}, 
S. Klein $^{\, b,}$\footnote{{\bf e-mail}: sebastian.klein@desy.de}, 
S. Moch$^{\, b,}$\footnote{{\bf e-mail}: sven-olaf.moch@desy.de} \\
\vspace{0.5cm}
\normalsize
{\it $^a$Institute for High Energy Physics \\
\vspace{0.1cm}
142281 Protvino, Moscow Region, Russia}\\
\vspace{0.5cm}   
{\it $^b$Deutsches Elektronensynchrotron DESY \\
\vspace{0.1cm}
Platanenallee 6, D--15738 Zeuthen, Germany}\\
\vspace{0.5cm}
\large {\bf Abstract}
\vspace{-0.2cm}
\end{center}
\noindent {\small
We determine the parton distribution functions (PDFs) in a next-to-next-to-leading order 
(NNLO) QCD-analysis of the  inclusive neutral-current deep-inelastic-scattering (DIS) world 
data combined with the neutri{-}no-nucleon DIS di-muon data and the fixed-target Drell-Yan data. 
The PDF-evolution is performed in the $N_f = 3$ fixed-flavor scheme and supplementary 
sets of PDFs in the 4- and 5-flavor schemes are derived from the results in 
the 3-flavor scheme using matching conditions. 
The charm-quark DIS contribution is
calculated in a general-mass variable-flavor-number (GMVFN) scheme 
interpolating between the zero-mass 4-flavor scheme at asymptotically large values of
momentum transfer $Q^2$ and the 3-flavor scheme at the value of $Q^2 = m_c^2$ in a prescription of 
Buza-Matiounine-Smith-van Neerven (BMSN). 
The results 
in the GMVFN scheme are compared with those of the fixed-flavor scheme and other 
prescriptions used in global fits of PDFs. The strong coupling constant is measured 
at an accuracy of $\approx 1.5\%$. We obtain at NNLO $\alpha_s(M_Z^2) = 0.1135 \pm 
0.0014$ in the fixed-flavor scheme and $\alpha_s(M_Z^2) = 0.1129 \pm 0.0014$ 
applying the BMSN prescription. The implications for important standard candle and hard 
scattering processes at hadron colliders are illustrated. Predictions for cross sections 
of $W^{\pm}$- and $Z$-boson, the top-quark pair- and Higgs-boson production at the 
Tevatron and the LHC based on the 5-flavor PDFs of the present analysis are provided.
}
\vspace{1.5cm}
\end{titlepage}

\setcounter{footnote}{0}
%%%%%%%%%%%%%%%%%%%%%%%%%%%%%%%%%%%%%%%%%%%%%%%%%%%%%%%%%%%%%%%%%%%%%%%%%%
\section{Introduction}
\label{sec1}
%%%%%%%%%%%%%%%%%%%%%%%%%%%%%%%%%%%%%%%%%%%%%%%%%%%%%%%%%%%%%%%%%%%%%%%%%%
\setcounter{table}{0}

\vspace{1mm}\noindent
For many hard processes at high energies heavy flavor production forms 
a significant part of the scattering cross section. As it is well known, the 
scaling violations are different in the massive and massless cases. 
Therefore, in all precision measurements, a detailed treatment of the
heavy flavor contributions is required. This applies, in particular to the
extraction of the twist-2 parton distribution functions (PDFs) in 
deep-inelastic scattering (DIS). In this process $O(25 \%)$ of
the inclusive cross section in the range of small values of $x$ is due 
to the production of charm-quarks as measured by the HERA experiments H1 and 
ZEUS~\cite{H1,ZEUS}. To perform a consistent QCD-analysis of the DIS 
world data and other hard scattering data, a next-to-next-to-leading order (NNLO) analysis 
is required, which 
includes the 3-loop anomalous dimensions \cite{Moch:2004pa} and the 
corresponding Wilson coefficients \cite{Zijlstra:1992qd}, in particular those for 
the heavy flavor contributions. The latter are known at leading order (LO) \cite{LO,Shifman:1977yb} 
and next-to-leading order (NLO) \cite{NLO}.
In the present paper we restrict the analysis to 
the NLO heavy flavor corrections. Very recently a series of Mellin moments 
at NNLO has 
been calculated in Ref.~\cite{Bierenbaum:2009mv} for the heavy flavor Wilson coefficients 
of the structure function $F_2$, in the region $Q^2 \gsim 10 \cdot m_h^2$, 
where $m_h$ is the heavy quark mass and $Q^2$ is the momentum transfer squared. 
Because of the large heavy flavor contribution to 
$F_2$, its correct description is essential in precision measurements 
of the strong coupling constant $\alpha_s$ and of the PDFs.

At asymptotically large values of $Q^2$, the heavy flavor contributions rise like 
$\alpha_s(Q^2) \ln(Q^2/m_h^2)$.
Despite the suppression due to the relatively small value of $\alpha_s$
at large scales, these terms might dominate and therefore their resummation 
is necessary~\cite{Shifman:1977yb}. It can be easily performed 
through the renormalization group equations for mass factorization for the 
process independent contributions defining the so-called variable-flavor-number (VFN) 
scheme. Thereby heavy quark PDFs are introduced, as e.g. suggested 
in Ref.~\cite{Aivazis:1993pi}. A VFN scheme has to be used in global fits of hadron collider data if the cross sections of the
corresponding processes are not available in the 3-flavor scheme. 
However, since VFN schemes are only applicable at asymptotically  
large momentum transfers, one has to find a description suitable for lower
virtualities $Q^2$, which matches with the 
3-flavor scheme at the scale $Q^2=m_h^2$, cf. Ref.~\cite{Jung:2009eq}. 

At the same time, 
the resummed large logarithms occur in the higher order corrections. 
In the NLO corrections to the massive electro-production coefficient
functions~\cite{NLO} the terms up to $\alpha_s^2(Q^2)\ln^2(Q^2/m_h^2)$
are manifest. Therefore the resummation of the remaining large logarithms 
is much less important as compared 
to the LO case. Furthermore, in most of the kinematic domain of the 
DIS experiments the impact of the resummation is 
insignificant~\cite{Gluck:1993dpa}. Eventually, the relevance of the resummation
is defined by the precision of the analyzed data and has to be checked 
in the respective cases. In this paper we study the impact of the heavy flavor 
corrections  
on the PDFs extracted from global fits including the most 
recent neutral-current DIS data. We apply the results of the QCD-analysis to 
main NNLO hard scattering cross sections, as the $W/Z$-gauge boson, 
top-quark pair and Higgs-boson production at hadron colliders.

The paper is organized as follows. In Section 2 we outline the theoretical formalism 
which describes the heavy quark contributions to DIS structure functions and
the formulation of VFN schemes, 
cf. Refs.~\cite{Buza:1996wv,Bierenbaum:2009zt,Bierenbaum:2009mv}. A phenomenological  
comparison of the fixed-flavor number (FFN) scheme and different VFN schemes is 
performed in Section~3 and 4. In Section 4 we present the results of an NNLO PDF-fit 
to the 
DIS world data, the fixed-target Drell-Yan- and di-muon data in different 
schemes using correlated errors to determine the PDF-parameters and 
$\alpha_s(M_Z^2)$. Precision predictions of PDFs are 
very essential for all measurements at hadron colliders 
\cite{HERALHC}.
Section 5 describes the 3-, 4-, and 5- flavor PDFs generated from the results
of our fit and applications to hadron collider phenomenology, such
as the scattering cross section of 
$W^{\pm}$- and $Z$-boson production, the top-quark pair and Higgs-boson
cross sections based on the 5-flavor PDFs obtained in the present analysis.
Section~6 contains the conclusions.

%%%%%%%%%%%%%%%%%%%%%%%%%%%%%%%%%%%%%%%%%%%%%%%%%%%%%%%%%%%%%%%%%%%%%%%%%%
\section{Heavy Quark Contributions: Theoretical Framework}
\label{sec2}
%%%%%%%%%%%%%%%%%%%%%%%%%%%%%%%%%%%%%%%%%%%%%%%%%%%%%%%%%%%%%%%%%%%%%%%%%%

\vspace{1mm}\noindent
In inclusive DIS, heavy quarks contribute to the final state if we consider extrinsic heavy 
flavor production only~\footnote{Potential contributions due to intrinsic charm were limited 
to be less than 1 \% in Ref.~\cite{INCHARM}.}. In fixed-order calculations of the
inclusive heavy flavor cross sections in the fixed flavor number scheme (FFNS) for
$N_f$ light quarks, one obtains the following representation for the DIS structure functions
to NLO in case of single photon exchange~\cite{LO,NLO,Bierenbaum:2009mv} 
%--------------------------------------------------------------------------
\begin{eqnarray}
\label{eqFimain}
F_i^{h, \rm exact}(N_f,x,Q^2) &=& \nonumber \\
&& 
\hspace*{-2.0cm}
\int_x^{x_{\rm max}}dz\left\{ 
e_h^2\left[
H_{g,i}(z,Q^2,m_h^2,\mu^2)\frac{x}{z}G\left(N_f,\frac{x}{z},\mu^2\right)
+H^{\rm PS}_{q,i}(z,Q^2,m_h^2,\mu^2)\frac{x}{z}\Sigma 
\left(N_f,\frac{x}{z},\mu^2\right)
\right] \right.
\nonumber \\ 
&& \left.  \hspace*{1.3cm} 
\hspace*{-2cm}
+ \sum_{k=1}^{N_l} e_k^2 
L_{g,i}(z,Q^2,m_h^2,\mu^2)\frac{x}{z}G\left(N_f,\frac{x}{z},\mu^2\right) 
+L^{\rm NS}_{q,i}(z,Q^2,m_h^2,\mu^2)\frac{x}{z}
f\left(N_f,\frac{x}{z},\mu^2\right) 
\right\}~,
\nonumber\\
\label{eqn:ffn}
\end{eqnarray}
%--------------------------------------------------------------------------
where $i = 2,L$. The functions $H_{g(q),i}$ and $L_{g(q),i}$ denote the massive Wilson 
coefficients with the photon coupling to the heavy ($H$) or
a light ($L$) quark line, respectively, $x = Q^2/(2 p.q)$ is the  Bjorken scaling variable,
with $q$ the 4--momentum transfer, $p$ the nucleon momentum, $Q^2 = - q^2$;
$x_{\rm max} = Q^2/(Q^2+4m_h^2)$ is production threshold;
$e_h$ is the charge of the heavy quark, with 
$h=c,b$. We introduced a second symbol for the number of the light flavors, 
$N_l$, which counts the number of the light quark anti-quark final state pairs associated to
the Wilson coefficients $L_{g,i}$. 
The flavor singlet and non-singlet distributions are given by
%--------------------------------------------------------------------------
\begin{eqnarray}
\Sigma \left(N_f,{x},\mu^2\right) &=&
\sum_{k=1}^{N_f}\left[ q_k\left(N_f,{x},\mu^2\right)
+\bar{q}_k\left(N_f,{x},\mu^2\right)\right], 
\label{eqn:sfffn1}
\\
f\left(N_f,{x},\mu^2\right)&=&\sum_{k = 1}^{N_f} 
e_k^2 \left[q_k\left(N_f,{x},\mu^2\right)
+\bar{q}_k\left(N_f,{x},\mu^2\right)\right]~,
\end{eqnarray}
%--------------------------------------------------------------------------
where $q_k$, $\bar{q}_k$ and $G$ are the light quark, anti-quark and gluon distributions.
Here and in the following we identify the factorization and renormalization scales
by $\mu = \mu_F = \mu_R$.
In open heavy flavor production one usually chooses $\mu^2 = {Q^2+4m_h^2}$,
while for the inclusive structure functions 
one sets $\mu^2 = Q^2$.

The massive Wilson coefficients in  Eq.~(\ref{eqFimain})
are available in analytic form at LO \cite{LO} and in semi-analytic form at NLO 
\cite{NLO}~\footnote{A fast implementation in Mellin space is given in 
Ref.~\cite{AB}.}. 
For $Q^2/m_h^2 \gg 1$ they were given in analytic form to NLO in 
Refs.~\cite{Buza:1995ie,Buza:1996wv,HQ2007} and in \cite{BFNK,Bierenbaum:2009mv} to NNLO 
for $F_L$ and $F_2$.
The NNLO contributions to $F_2$ 
are not yet fully available as general expressions in $x$ or the Mellin variable 
$N$, since for one part, only a series of 
Mellin moments at fixed integer values of $N$ has been calculated so far~\cite{Bierenbaum:2009mv}.
In the limit $Q^2 \gg m^2_h$, the integration in  Eq.~(\ref{eqFimain}) extends to $x_{\rm max} = 1$ 
and 
additional soft- and virtual terms contribute to the cross section according to the 
Kinoshita-Lee-Nauenberg theorem, cf. e.g.~\cite{Buza:1995ie}. 

In Ref.~\cite{NLO} the effects due to heavy quark loops in external gluon lines
were absorbed for the heavy flavor Wilson coefficients into the strong coupling 
constant to NLO, which is then to be taken in the corresponding momentum subtraction scheme
in  Ref.~\cite{Bierenbaum:2009mv}. 
The necessary changes for $\alpha_s$ in the $\overline{\rm MS}$-scheme are discussed 
in Refs.~\cite{Buza:1996wv,Bierenbaum:2009mv,Bierenbaum:2009zt}. 
In the present paper, we will include the NLO contributions for 
$F_L$ and $F_2$ with $\alpha_s$ in the $\overline{\rm MS}$-scheme, 
cf. Ref.~\cite{Bierenbaum:2009mv}.
The choice of a {\sf MOM} scheme always forms an intermediate step, since it applies 
to the heavy degrees of freedom only. The structure functions also contain the light
flavor PDFs and massless Wilson coefficients, the scaling violations 
of which are governed by $\alpha_s^{\overline{\rm MS}}$ only. Also, one cannot 
choose a scheme, which introduces heavy quark mass effects in the strong coupling constant 
below any heavy flavor threshold.

%---------------------------------------------------------------------------------
In the asymptotic region  $Q^2\gg m_h^2$ the Wilson 
coefficients $L_{g(q),2}$ and $H_{g(q),2}$ for
the heavy flavor structure function $F_2^{h, \rm exact}$ of 
Eq.~(\ref{eqFimain})
can be expressed in terms of the massive 
operator matrix elements
$A_{ij}$ and the massless Wilson coefficients $C_{k,2}$. The former are given by
\begin{eqnarray}
\label{eqAQg1}
A_{ij}\left(N_f,z, \frac{m_h^2}{\mu^2}\right) &=& \delta_{ij} 
+ \sum_{n=1}^\infty
a_s^n(N_f,\mu^2)
A_{ij}^{(n)}\left(N_f,z, \frac{m_h^2}{\mu^2}\right),~~~~~i,j~\in~\{h,q,g\};
\\
A_{ij}^{(1)}\left(z,\frac{m_h^2}{\mu^2}\right)
&=&
a_{ij}^{(1,1)}(z)\ln\left(\frac{\mu^2}{m_h^2}\right)
+ a_{ij}^{(1,0)}(z)\, ,\\
A_{ij}^{(2)}\left (z,\frac{m_h^2}{\mu^2}\right )&=&
a_{ij}^{(2,2)}(z)\ln^2\left(\frac{\mu^2}{m_h^2}\right) +
a_{ij}^{(2,1)}(z)\ln\left(\frac{\mu^2}{m_h^2}\right)  +
a_{ij}^{(2,0)}(z)~,
\end{eqnarray}
cf. 
Refs.~\cite{Buza:1995ie,Buza:1996wv,HQ2007,Bierenbaum:2008yu,Bierenbaum:2009zt}.
To NLO the massive OMEs do not depend on $N_f$.
The massless Wilson coefficients  
for the structure function $F_2$ are given by
\begin{eqnarray}
C_{k,2}\left(N_f,z,\frac{Q^2}{\mu^2}\right) &=& 
\sum_{n=0}^\infty a_s^n(N_f,\mu^2) 
C_{k,2}^{(n)}\left(N_f,z,\frac{Q^2}{\mu^2}\right),~~~~~k=q,g,
\label{eqn:c2}
\end{eqnarray}
cf. Refs.~\cite{Zijlstra:1992qd,Vermaseren:2005qc}. In case of $C_{q,2}$ we decompose
the Wilson coefficients into flavor non-singlet (NS) and pure-singlet (PS) contributions 
$C^{\rm NS}_{q,2}$ and $C^{\rm PS}_{q,2}$.
We use the strong coupling constant in the notation
$a_s(N_f,\mu^2) = \alpha_s(N_f,\mu^2)/(4 \pi)$. At the different heavy flavor thresholds
$\mu^2 = m_h^2,~~h = c, b,$ matching conditions are employed to $a_s(\mu^2)$, cf. e.g.
Ref.~\cite{BETH}.

Up to $O(\alpha_s^2)$ the asymptotic expressions for the 
heavy flavor coefficients $L_{g(q),2}$ and $H_{g(q),2}$ 
read~\cite{Buza:1995ie,Bierenbaum:2009mv}
%---------------------------------------------------------------------------------
\begin{eqnarray}
\label{W1}
L_{q,2}^{\rm asymp, \rm NS} &=& a_s^2(N_f) \left\{A_{qq,h}^{(2),\rm NS}
+ \left[C_{q,2}^{(2),\rm NS}(N_f+1) - C_{q,2}^{(2),\rm NS}(N_f) \right] 
\right\} 
\, ,\\
L_{g,2}^{\rm asymp} &=& a_s^2(N_f) A_{gg,h}^{(1)}
\otimes~
\frac{1}{N_f} C_{g,2}^{(1)}(N_f) 
\, ,\\
H_{q,2}^{\rm asymp, \rm PS} &=& a_s^2(N_f) \left[A_{hq}^{(2),\rm PS} + 
\frac{1}{N_f} C_{q,2}^{(2),\rm PS}(N_f)\right]
\, ,\\
\label{W2}
H_{g,2}^{\rm asymp} &=& a_s(N_f) \left[A_{hg}^{(1)} 
                                     +  \frac{1}{N_f} C_{g,2}^{(1)}(N_f)\right]
+ a_s^2(N_f) \Biggl\{A_{hg}^{(2)} 
+A_{hg}^{(1)} \otimes {C}_{q,2}^{(1),\rm NS} \nonumber\\
&& \hspace{11mm}
+A_{gg,h}^{(1)} \otimes  \frac{1}{N_f} C_{g,2}^{(1)}(N_f)
+  \frac{1}{N_f}C_{g,2}^{(2)}(N_f)\Biggr\}~. 
\end{eqnarray}
%-------------------------------------------------------------------------------
The symbol $\otimes$ denotes the Mellin convolution 
%---------------------------------------------------------------------------------
\begin{eqnarray}
[A \otimes B](z) = \int_z^1 \frac{dy}{y} A(y) B\left(\frac{z}{y}\right)~
\end{eqnarray}
and all arguments except of $N_f$ are omitted for brevity. Note that nearly identical graphs 
contribute 
to $L_{g,2}^{\rm asymp}$ and the second last term of $H_{g,2}^{\rm asymp}$. These are accounted
for in different classes due to the final
state fermion pair, which consists of the light quarks in the first case and the heavy quark
in the second case. Therefore  we introduced
$N_l$ as a second label for the number of light flavors in the final state, 
cf.~Eq.~(\ref{eqFimain}).

The OMEs enter in the matching conditions for the PDFs in the $N_f$-flavor
scheme with the ones for $(N_f+1)$ massless flavors~\cite{Buza:1996wv} which are
implied by the renormalization group equations. 
In particular, the NNLO heavy-quark distribution
in the $(N_f+1)$-flavor scheme at  $O(a_s^2)$ 
reads
%-----------------------------------------------------------------------------
\begin{eqnarray}
  h^{(1)}(x,\mu^2)+\bar{h}^{(1)}(x,\mu^2) &=& a_s(N_f+1,\mu^2)
    \Biggl[
          A_{hg}^{(1)}\Biggl(\frac{m_h^2}{\mu^2}\Biggr)
          \otimes G^{(2)}\Bigl(N_f,\mu^2\Bigr)
    \Biggr](x),
%--
\label{fQQB}
%\nonumber\\
\end{eqnarray}
%------------------------------------------------------------------------------------------
\begin{eqnarray}
\hspace*{1cm}
h^{(2)}(x,\mu^2)+   \bar{h}^{(2)}(x,\mu^2)= h^{(1)}(x,\mu^2)+   \bar{h}^{(1)}(x,\mu^2) \nonumber\\ && 
\hspace*{-9cm}
+a_s^2(N_f+1,\mu^2)
\left\{ \left[ A_{hg}^{(2)}\left(\frac{m_h^2}{\mu^2}\right) \otimes 
G^{(2)}\left(N_f,\mu^2\right)\right](x)
+\left[A_{hq}^{(2),\rm PS}\left(\frac{m_h^2}{\mu^2}\right)
\otimes
\Sigma^{(2)}\left(N_f,\mu^2\right)\right](x)\right\}, 
\label{eqn:hqpdf}
\end{eqnarray}
%-----------------------------------------------------------------------------
where $G^{(2)}$ and $\Sigma^{(2)}$ are the gluon and flavor singlet
distributions, respectively, evolved at NNLO. 
Likewise, one obtains for the gluon, flavor non-singlet and singlet distributions in the
$N_f+1$-flavor scheme up to $O(a_s^2)$
%-----------------------------------------------------------------------------
\begin{eqnarray}
G^{(2)}(N_f+1,x,\mu^2) &=&G^{(2)}(N_f,x,\mu^2) + a_s(N_f+1, \mu^2)
\left[A_{gg,h}^{(1)}\left(\frac{m^2_h}{\mu^2}\right) \otimes
G^{(2)}\left(N_f,\mu^2\right)\right](x)
\nonumber \\ &&
+ a_s^2(N_f+1,\mu^2) \Biggl\{\left[
A_{gg,h}^{(2)}\left(\frac{m^2_h}{\mu^2}\right) \otimes
G^{(2)}\left(N_f,\mu^2\right)\right](x)
\nonumber\\ &&
\hspace*{2.5cm} +
\left[A_{gq}^{(2)}\left(\frac{m^2_h}{\mu^2}\right) \otimes
\Sigma^{(2)}\left(N_f,\mu^2\right)\right](x)\Biggr\}~,
\label{HPDF2}
\\
\Sigma^{(2)}(N_f+1,x,\mu^2)&=&\Sigma^{(2)}(N_f,x,\mu^2)
+ a_s(N_f+1,\mu^2) \left[A_{hg}^{(1)}\left(\frac{m^2_h}{\mu^2}\right) \otimes
G^{(2)}\left(N_f,\mu^2\right)\right](x)
\nonumber
\end{eqnarray}
\begin{eqnarray}
&& \hspace*{2.3cm}
+
a_s^2(N_f+1,\mu^2)\left[
 A_{qq,h}^{(2),\rm NS}\left(\frac{m^2_h}{\mu^2}\right)
+A_{hq}^{(2),\rm PS}\left(\frac{m^2_h}{\mu^2} \right) \right]
\otimes \Sigma^{(2)}\left(N_f,\mu^2\right)(x)
\nonumber\\
&&
\hspace*{2.3cm}
+
a_s^2(N_f+1,\mu^2) \left[A_{hg}^{(2)}\left(\frac{m^2_h}{\mu^2}\right) \otimes
G^{(2)}\left(N_f,\mu^2\right)\right](x)~,
\label{eqn:sigma4pdf}
\end{eqnarray}
%-----------------------------------------------------------------------------
and the light quark and anti-quark distributions are given by
%-----------------------------------------------------------------------------
\begin{eqnarray}
\hspace*{-5mm}
q_k^{(2)}(N_f+1,x,\mu^2) +
\overline{q}_k^{(2)}(N_f+1,x,\mu^2) && \nonumber\\ &&
\hspace*{-4.3cm} =
\left[1 + a_s^2(N_f+1, \mu^2) A_{qq,h}^{(2), \rm NS}\left(\frac{m_h^2}{\mu^2}\right)\right]
\otimes \left[
q_k^{(2)}(N_f,x,\mu^2) +
\overline{q}_k^{(2)}(N_f,x,\mu^2)\right]~.
\label{HPDF4}
\end{eqnarray}
%-----------------------------------------------------------------------------
These distributions obey momentum conservation
%-----------------------------------------------------------------------------
\begin{eqnarray}
\small
1 &=& \int_0^1 dx~x~\left[ G(N_f,\mu^2,x) + \Sigma(N_f,\mu^2,x) \right] \nonumber\\
\small
&& \hspace*{-12mm}
  =
\int_0^1 dx~x~ \left\{ G(N_f+1,\mu^2,x) +  
\sum_{k=1}^{N_f}\left[ q_k\left(N_f+1,{x},\mu^2\right)
+\bar{q}_k\left(N_f+1,{x},\mu^2\right)\right]
+h^{(2)}(x,\mu^2)+   \bar{h}^{(2)}(x,\mu^2) \right\}.
\nonumber\\
\normalsize
\end{eqnarray}
%----------------------------------------------------------------------------
Since the OMEs are process independent quantities this property is maintained by the  
{$(N_f+1)$-flavor} PDFs.
One may apply these PDFs in a hard scattering process for large enough scales
$\mu_F^2\gg m_h^2$, where the power corrections are negligible.
In particular, the heavy flavor structure function $F_2$ 
is defined in the $(N_f+1)$-flavor scheme as the convolution 
of the $(N_f+1)$-flavor PDFs with the massless 
Wilson coefficients $C_{q(g),2}$. This representation is the so-called 
zero mass VFN (ZMVFN) scheme expression, which is applicable 
only in the asymptotic region.

The heavy flavor part of $F_2$ in the region $Q^2 \gg m_h^2$ for $N_f+1$ flavors
up to $O(\alpha_s^2)$ is given by
\begin{eqnarray}
  F_2^{h,\rm ZMVFN}(N_f+1,x,Q^2)&=&
    xe_h^2\Biggl\{
            h^{(2)}(x,\mu^2)+ \bar{h}^{(2)}(x,\mu^2)
\nonumber 
\\ 
%\end{eqnarray}\begin{eqnarray}
&& \hspace{-35mm}
            +a_{\rm s}(N_f+1,\mu^2)
              \Biggl[
\frac{1}{N_f}C_{g,2}^{(1)}\Biggl(N_f,\frac{Q^2}{\mu^2}\Biggr)
                  \otimes G^{(2)}\Bigl(N_f,\mu^2\Bigr)
              \Biggr](x)
\nonumber \\ && \hspace{-35mm}
            +a^2_{\rm s}(N_f+1,\mu^2)
              \Biggl[
                  A^{(1)}_{gg,h}\Biggl(\frac{m_h^2}{\mu^2}\Biggr)\otimes
                  \frac{1}{N_f} 
C_{g,2}^{(1)}\Biggl(N_f,\frac{Q^2}{\mu^2}\Biggr)
                  \otimes G^{(2)}\Bigl(N_f,\mu^2\Bigr)
              \Biggr](x)
\nonumber \\ && \hspace{-35mm}
            +a_{\rm s}(N_f+1,\mu^2)
              \Biggl[
                  C_{q,2}^{(1),{\rm 
NS}}\Biggl(\frac{Q^2}{\mu^2}\Biggr)\otimes
                  \Bigl[
h^{(1)}\Bigl(\mu^2\Bigr)+\bar{h}^{(1)}\Bigl(\mu^2\Bigr)
                  \Bigr]
              \Biggr](x)
\nonumber \\ %&& %\hspace{-20mm}
            && \hspace{-35mm} +\frac{1}{N_f} a^2_{\rm s}(N_f+1,\mu^2)
              \Biggl(
                    \Biggl[
                       C_{q,2}^{(2),{\rm 
PS}}\Biggl(N_f,\frac{Q^2}{\mu^2}\Biggr)
                          \otimes\Sigma^{(2)}\Bigl (N_f,\mu^2\Bigr)
                    \Biggr](x)
                   +\Biggl[
                          C_{g,2}^{(2)}\Biggl(N_f,\frac{Q^2}{\mu^2}\Biggr)
                          \otimes G^{(2)}\Bigl(N_f,\mu^2\Bigr)
                    \Biggr](x)
             \Biggr)
          \Biggr\}
\nonumber \\ && \hspace{-35mm}
            +x \frac{1}{N_f}\sum_{k=1}^{N_l} e_k^2  a^2_{\rm s}(N_f+1,\mu^2)
              \Biggl[
                  A^{(1)}_{gg,h}\Biggl(\frac{m_h^2}{\mu^2}\Biggr)\otimes
                  C_{g,2}^{(1)}\Biggl(N_f,\frac{Q^2}{\mu^2}\Biggr)
                  \otimes G^{(2)}\Bigl(N_f,\mu^2\Bigr)
              \Biggr](x)
\nonumber \\ && \hspace{-35mm}
   +x a^2_{\rm s}(N_f+1,\mu^2)
          \Biggl[
               \Biggl(
                     A_{qq,h}^{(2),{\rm NS}}\Biggl(\frac{m_h^2}{\mu^2}\Biggr)
                    +C_{q,2}^{(2),{\rm NS}}\Biggl(N_f+1,\frac{Q^2}{\mu^2}\Biggr)
                    -C_{q,2}^{(2),{\rm NS}}\Biggl(N_f,\frac{Q^2}{\mu^2}\Biggr)
               \Biggr)\otimes f\Bigl(N_f,\mu^2\Bigr)\Biggr](x) \, .
     \nonumber \\
  \label{eqn:zmvfn}
\end{eqnarray}

At this point we would briefly like to comment on the 
longitudinal structure function $F_L$. As a matter of fact, 
the above concept of a ZMVFN scheme cannot be directly 
applied to the heavy flavor component of $F_L$ even 
in the asymptotic regime of $Q^2\gg m_h^2$; 
e.g. at $O(\alpha_s)$, similarly to Eq. (\ref{eqn:zmvfn}), one obtains
%----------------------------------------------------------------------------
\begin{equation}
F_L^{h,\rm asymp}(N_f+1,x,Q^2) = 
a_s(N_f+1,\mu^2) e_h^2 \left[C_{g,L}^{(1)}\left(\frac{Q^2}{\mu^2}\right)
\otimes G(N_f,\mu^2)\right](x)~,
\end{equation}
%----------------------------------------------------------------------------
Here, the gluon density is convoluted with the LO gluon Wilson coefficient
$C_{g,L}^{(1)}$ but not a splitting function, 
because unlike the case of $F_2$ no collinear logarithm emerges.
The example illustrates that a detailed renormalization group analysis is a
necessary prerequisite to the use of heavy quark densities 
even in the asymptotic region, cf. Ref.~\cite{Buza:1996wv}.

%%%%%%%%%%%%%%%%%%%%%%%%%%%%%%%%%%%%%%%%%%%%%%%%%%%%%%%%%%%%%%%%%%%%%%%%%
\section{Comparison of the 3- and the 4-Flavor Schemes}
\label{sec3}
%%%%%%%%%%%%%%%%%%%%%%%%%%%%%%%%%%%%%%%%%%%%%%%%%%%%%%%%%%%%%%%%%%%%%%%%%

\vspace{1mm}\noindent
%----------------------------------------------------------------------------
At $O(\alpha_s^l)$ the universal contribution (referring to the massive OMEs
only) to the heavy flavor singlet contribution to 
$F_2$ 
is given by 
%--------------------------------------------------------------------------
\begin{eqnarray}
\widehat{F}_2^{h,(l)}(N_f=4,x,Q^2) = e_h^2 x
[h^{(l)}(x,\mu^2)+\bar{h}^{(l)}(x,\mu^2)]~,~~~~l = 1,2~.
\end{eqnarray}
%--------------------------------------------------------------------------
\begin{figure}[h]
  \begin{center}
    \includegraphics[width=11.0cm]{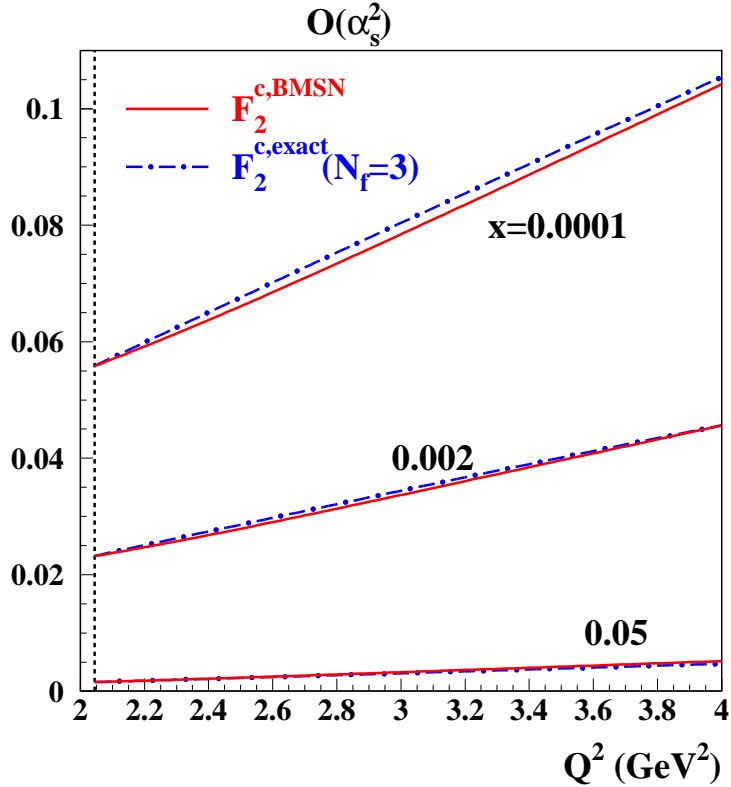}
    \vspace*{-5mm}
    \caption[]{ \small
     Matching of $F_2^{c, \rm BMSN}(N_f=4,x,Q^2)$ (solid lines) with 
     $F_2^{c,\rm exact}(N_f=3,x,Q^2)$ (dash-dotted lines) at 
     small $Q^2$ in $O(\alpha^2_s)$. 
     The vertical line denotes the position of the charm-quark mass
     $m_c=1.43~{\rm GeV}$.
      \label{fig:match}
    }
  \end{center}
\end{figure}
It vanishes for $\widehat{F}_2^{h,(1)}$ at $\mu^2 = m_h^2$, 
since $a_{hg}^{(1,0)} = 0$, cf. Eq.~(\ref{eqAQg1}), and 
it is negative for $\mu^2<m_h^2$. 
However, the 1st order heavy quark contribution to the structure function $F_2$ 
is positive, since the $\mu^2$ dependence is canceled by a corresponding 
logarithm $\propto \ln(Q^2/\mu^2)$ in the massless Wilson 
coefficient $C_{g,2}^{(1)}$ in Eq.(\ref{eqn:zmvfn}).
Despite that in the 3-flavor scheme the heavy quark 
contribution to the DIS structure functions also falls at small $Q^2$, it is 
present
down to the photo-production limit. At $O(\alpha_s^2)$ the agreement 
between the two schemes at low $Q^2$ is even worse since the term 
$a_{hg}^{(2,0)}$ is negative which implies $\widehat{F}_i^{h,(2)}(N_f=4,x,Q^2) < 0$
because the gluon contribution dominates over the pure-singlet part numerically. 
The impact of the large-log resummation is negligible at small scales 
$Q^2$ and any reasonable scheme must reproduce the 3-flavor scheme. 
%---------------------------------------------------------------------------------
\begin{figure}[htb]
  \begin{center}
    \includegraphics[width=16.5cm]{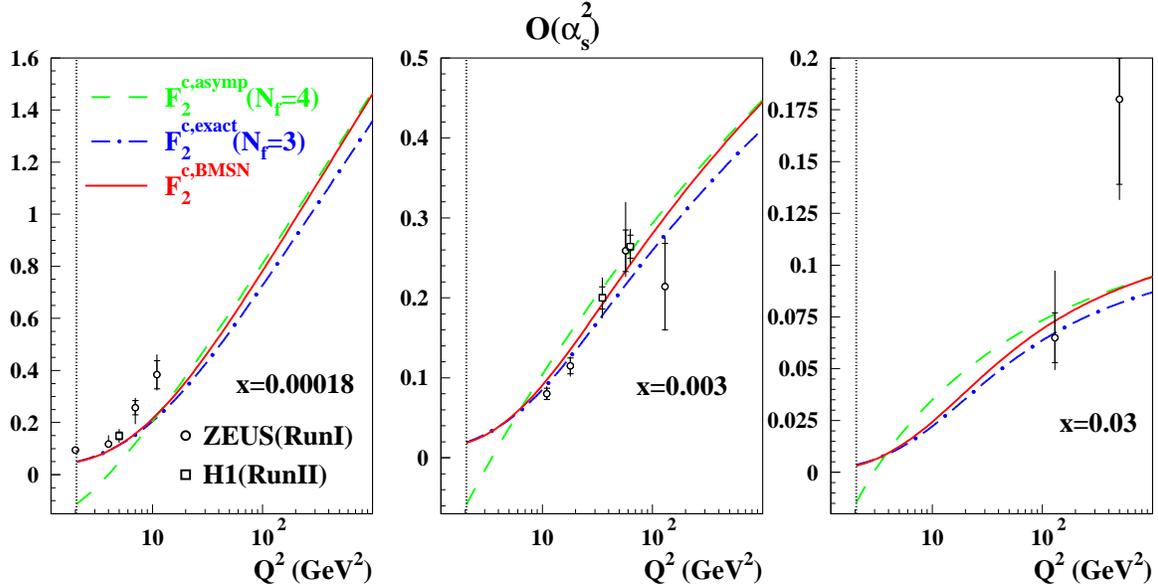}
    \vspace*{-5mm}
    \caption[]{ \small Comparison of $F_2^c$ in different 
    schemes to  H1- and ZEUS-data. Solid lines: GMVFN scheme in the 
    BMSN prescription, 
    dash-dotted lines: 3-flavor scheme, 
    dashed lines: 4-flavor scheme. The vertical dotted 
    line denotes the  position of the charm-quark mass $m_c=1.43~{\rm GeV}$.
      \label{fig:f2c}
    }
  \end{center}
\end{figure}
%---------------------------------------------------------------------------------
Therefore, at low values of $Q^2$ the ZMVFN scheme is not applicable. It 
has 
to be modified according to practical purposes. VFN schemes with such 
modifications are called general-mass variable-flavor-number (GMVFN) schemes, in contrast to the 
ZMVFN scheme. 
A particular form of the GMVFN scheme cannot be derived from first principles in 
an unique way, but is subject to the corresponding prescription. Consistent
schemes have to obey renormalization group equations to not violate the
running of the coupling constant and masses, and to obey correct scale 
evolution. As 
a general requirement any 
such prescription should provide a continuous transition from the 3-flavor 
scheme 
at low values of $\mu^2$ to the 4-flavor scheme at large scales. 

An early formulation of a GMVFN scheme by Aivazis-Collins-Olness-Tung 
(ACOT)~\cite{Aivazis:1993pi} does not allow a smooth matching with the  
3-flavor scheme at small scales $Q^2$. In the ACOT scheme, the slope in $Q^2$ 
turns out to be too large. Later the so-called Thorne-Roberts (TR) scheme 
overcoming this shortcoming was suggested~\cite{Thorne:1997uu}. 
However, beyond NLO this scheme is very involved and its numerical implementation 
is problematic~\cite{Thorne:1997ga}. Recently, the early ACOT prescription 
has been modified in order to improve the behavior at low values of 
$Q^2$~\cite{Tung:2001mv}. This modified description, the so-called ACOT($\chi$) scheme,
is used in particular at NNLO in Ref.~\cite{Thorne:2006qt}. Another 
GMVFN prescription,
which was suggested earlier by Buza-Matiounine-Smith-van Neerven
(BMSN)~\cite{Buza:1996wv} for $F_2^h$, is defined by 
%---------------------------------------------------------------------------------
\begin{equation}
F_2^{h,\rm BMSN}(N_f+1,x,Q^2) = F_2^{h,\rm exact}(N_f,x,Q^2)+F_2^{h,\rm 
ZMVFN}(N_f+1,x,Q^2)
                           -F_2^{h,\rm asymp}(N_f,x,Q^2)~,
\label{eqn:BMSN}
\end{equation}
%---------------------------------------------------------------------------------
with $N_f = 3$ for $h = c$.

Note that the difference of the last two terms in Eq.~(\ref{eqn:BMSN}) depends on $N_f$ through 
the strong coupling constant only, which is a specific feature up to NLO. 
For the choice of $\mu^2 = Q^2$ the asymptotic terms 
cancel at $Q^2 = m_h^2$  in Eq.~(\ref{eqn:BMSN}). 
In this limit $F_2^{h, \rm BMSN}(N_f=4)$ reproduces the result in the 3-flavor 
scheme.
Moreover,  $F_2^{h, \rm BMSN}(N_f = 4)$ matches with the 3-flavor scheme smoothly 
as shown in Figure~\ref{fig:match}. Minor kinks between 
$F_2^{h, \rm BMSN}(N_f=4)$ and $F_2^{h,\rm exact}(N_f=3)$ stem from the
matching of $\alpha_s(N_f,\mu^2)$ at $\mu^2 = m_h^2$. It
appears since 
the matching condition for $\alpha_s(N_f,\mu^2)$ 
does provide a continuous but not 
a smooth transition at the flavor thresholds.
The numerical impact 
of this kink is marginal in the analysis of the current data. 
At large $Q^2$ the asymptotic expression $F_2^{h,\rm asymp}(N_f=3)$ cancels the term 
$F_2^{h,\rm exact}(N_f=3)$ in Eq.~(\ref{eqn:BMSN}) 
and $F_2^{h, \rm BMSN}(N_f=4)$ 
reproduces the result in the ZMVFN scheme. The cancellation is not perfect due to 
the difference in the upper limit of integration in Eq.~(\ref{eqFimain}) and the 
expression for the ZMVFN scheme, 
which affects 
only the non-singlet Compton-type 
contribution given by the coefficient functions $L^{\rm NS}_{q,i}$. For the 3-flavor 
expression
of Eq.~(\ref{eqFimain}), this term rises as $\ln^3(Q^2/m_h^2)$ at large $Q^2$. 
In the asymptotic limit of Ref.~\cite{Buza:1995ie}
the corresponding singular contribution  
is washed out in $L_{q,2}^{\rm asymp,NS}$. As a result, there remains  
a contribution $\sim \ln^3(Q^2/m_h^2)$ in the difference of 
$F_2^{h,\rm exact}(N_f=3)$ 
and $F_2^{h,\rm asymp}(N_f=3)$. This mismatch is caused by a well-known 
soft and virtual term, which occurs in the
inclusive analysis for large arguments of the Wilson coefficient and is easily corrected, 
cf. Ref.~\cite{Buza:1995ie,Chuvakin:1999nx}.
On the other hand, the non-singlet contribution to heavy quark 
electro-production is numerically very small and the term 
$\sim \ln^3(Q^2/m_h^2)$ is apparent only at very large values of $Q^2$ and 
relatively large $x$. The accuracy of realistic data at this kinematics is rather 
poor and even for the definition  
of Eq.~(\ref{eqFimain}) the impact of the mismatch between 
$F_2^{h,\rm exact}(N_f=3)$ and 
$F_2^{h, \rm asymp}(N_f=3)$ turns out to be marginal in the data analysis.
%-------------------------------------------------------------------------------
\begin{figure}[htb]
  \begin{center}
    \includegraphics[width=14.0cm]{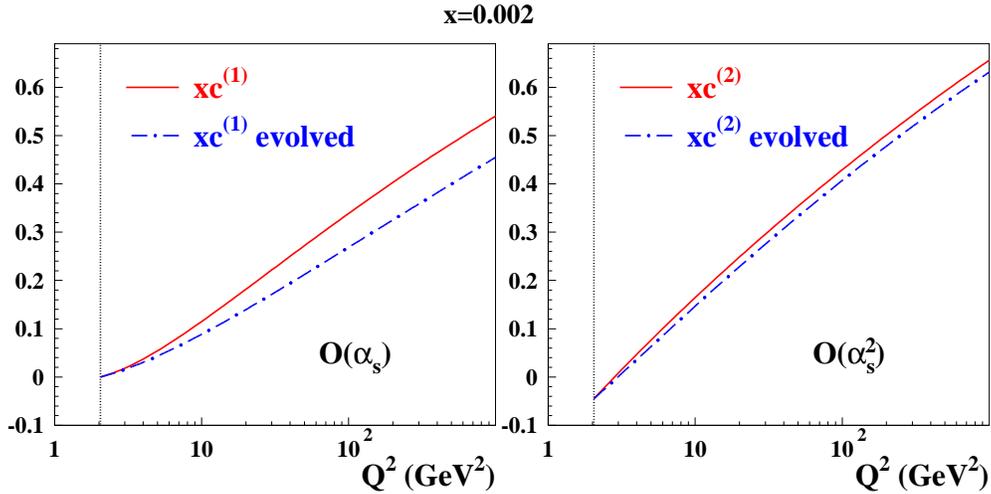}
    \vspace*{-5mm}
    \caption[]{ \small The $c$-quark distributions calculated using the 
    fixed-order relation Eqs.~(\ref{eqn:hqpdf},\ref{fQQB}) (solid lines) 
    compared to 
    the result in the 4-flavor scheme evolving from $m_c^2$ and using
    Eqs.~(\ref{eqn:hqpdf},\ref{fQQB}) as boundary condition (dash-dotted lines)
    at $O(\alpha_s)$ (left panel) and at $O(\alpha_s^2)$ (right panel). 
      \label{fig:evol}
    }
  \end{center}
\end{figure}
%-------------------------------------------------------------------------------
%-------------------------------------------------------------------------------
\begin{figure}[htb]
  \begin{center}
    \includegraphics[width=16.5cm]{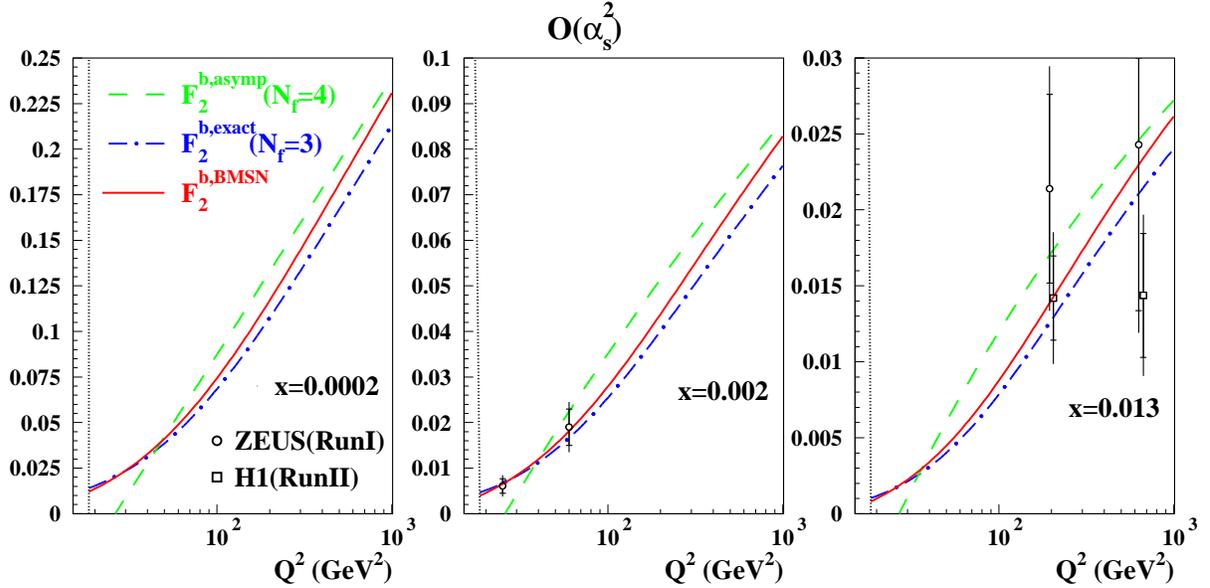}
    \vspace*{-5mm}
    \caption[]{ \small Comparison of predictions in different schemes 
              to ZEUS- and H1-data on $F_2^b(x,Q^2)$. The notations are the 
              same as in Figure~\ref{fig:f2c}. The vertical dotted line 
              marks the position of $m_b=4.3~{\rm GeV}$.
      \label{fig:f2b}
    }
  \end{center}
\end{figure}
%-------------------------------------------------------------------------------
%-------------------------------------------------------------------------------
\begin{figure}[htb]
  \begin{center}
    \includegraphics[width=16.5cm]{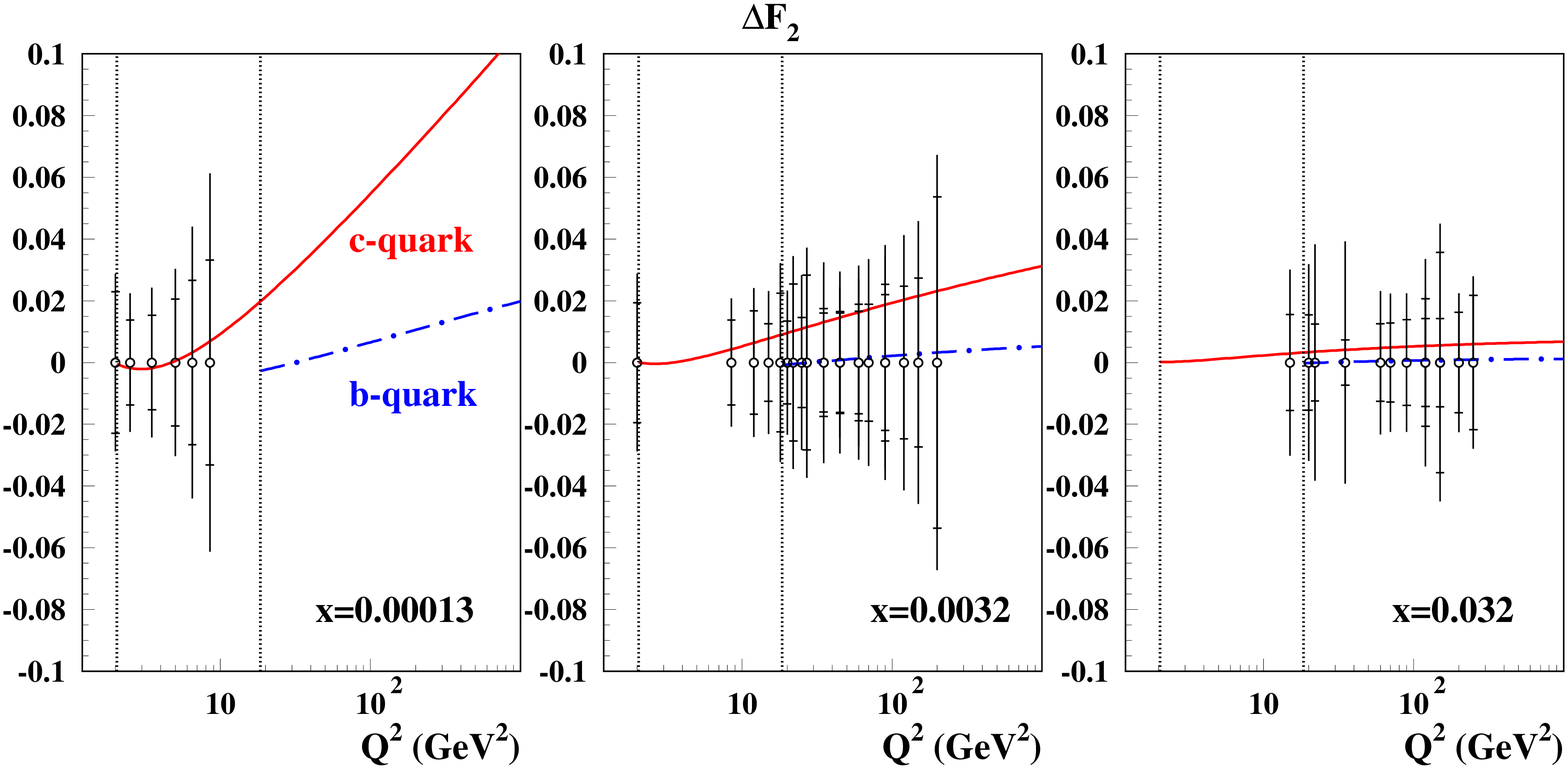}
    \vspace*{-5mm}
    \caption[]{ \small The errors in the inclusive structure function 
     $F_2(x,Q^2)$ measured by the H1 collaboration~\cite{h1incl} in comparison 
     with the impact of the heavy quark scheme variation on the QCD 
calculations 
     for $F_2(x,Q^2)$. Solid line: $c$-quark contribution, dash-dotted lines: 
     $b$-quark contribution. The vertical dots mark the positions of 
     $m_c=1.43~{\rm GeV}$ and $m_b=4.3~{\rm GeV}$, respectively. 
      \label{fig:f2incl}
    }
  \end{center}
\end{figure}
%-------------------------------------------------------------------------------

A representative set of the ZEUS- and H1-data~\cite{ZEUS,H1c} on 
$F_2^c$
is compared to $F_2^{c,\rm BMSN}(N_f=4)$, $F_2^{c,\rm exact}(N_f=3)$, and 
$F_2^{c,\rm ZMVFN}(N_f=4)$ in Figure~\ref{fig:f2c}. 
The 3-flavor PDFs used in this comparison are evolved starting from 
$m_c = 1.43~{\rm GeV}$ with the input given by the MRST2001 PDFs of 
Ref.\cite{Martin:2002aw}. 
Because of the kinematic constraints, at $x\sim 0.0001$ only the values 
of $Q^2\lesssim 10~{\rm GeV}^2$ are available in the data. At such low scales 
the calculation in the 3-flavor and BMSN scheme yield practically the same
results. At $x\sim 0.01$ the typical values of $Q^2$ are much bigger.
In this kinematic region, the BMSN scheme yields a larger contribution than 
obtained in the  3-flavor scheme. However, the uncertainties in the data are 
still quite large due to
limited statistics. The comparison of the calculations with the data is rather 
insensitive to the choice of scheme. The non-singlet term in Eq.~(\ref{eqFimain})
is not taken into account in the comparisons shown in Figure~\ref{fig:f2c}. 
Its impact is most significant at large values of $x$ and $Q^2$, 
but even in this 
case, it is much smaller than the data uncertainty.  For intermediate 
values 
of $x\sim 0.001$, a combination of these two cases is observed: 
at large $Q^2$ the uncertainties in the data do not allow to distinguish 
between both
schemes, while at small $Q^2$ the numerical difference between the 
3-flavor and the BMSN scheme calculations is very small. Summarizing, for the 
analysis of realistic data on $F_2^c$, the BMSN scheme is very similar 
to the 3-flavor scheme. This is not a particular feature of the 
BMSN prescription, 
since the difference  between 3- and 4-flavor schemes at large $Q^2$ 
is also smaller than the uncertainties in the available data 
and, once the smooth matching is provided, a GMVFN scheme must 
be close to the 3-flavor one at small $Q^2$. This conclusion is in 
agreement with the results of Ref.~\cite{Chuvakin:2000zj}. It derives from the 
fact that once the $O(\alpha_s^2)$ corrections are taken into account the 
need of a large-log resummation is thus greatly reduced, 
which is well known for a long time, cf. Ref.~\cite{Gluck:1993dpa}.  
In Figure~\ref{fig:evol} the
$c$-quark distribution defined in Eq.~(\ref{fQQB},\ref{eqn:hqpdf}) is compared to the
one evolved in the 4-flavor scheme starting from the scale of $m_c$ 
using Eqs.~(\ref{fQQB},\ref{eqn:hqpdf})
as boundary conditions. The former is derived from  fixed-order
perturbation theory, while for the latter resummation is performed
through the evolution equations. At $O(\alpha_s)$ the 
difference between these two approaches is significant indeed, however, at 
$O(\alpha_s^2)$ it is much smaller, and quite unimportant for  
realistic kinematics.

As evident from Figure~\ref{fig:f2c} the scheme choice cannot resolve 
observed discrepancies between data and the theoretical predictions to NLO.
Given the mass of the charm quark and the PDFs determined in inclusive 
analyses, higher order QCD corrections are needed. In particular, at small $x$ 
and $Q^2$, the partial $O(\alpha_s^3)$
corrections to the massive Wilson coefficient $H_{g,2}$ obtained through threshold 
resummation~\cite{Laenen:1998kp}
give a significant contribution to $F_2^c$ and 
greatly improve the agreement to the data~\cite{Alekhin:2008hc}.
In this kinematic region, the integral of 
Eq.~(\ref{eqFimain}) is mostly sensitive 
to the threshold of heavy quark production and the approximate form of 
$H_{g,2}$ derived in Ref.~\cite{Alekhin:2008hc} is sufficient.  
At large values of $Q^2$, the threshold approximation is inapplicable, and a 
complete NNLO calculation is required, cf. Ref.~\cite{Bierenbaum:2009mv}. 
For $b$-quark production the resummation 
effects are less important since the asymptotic region is scaled  
to bigger values of $Q^2$ and the data are less precise due to the smaller 
scattering cross section.
{This is illustrated by a comparison of the ZEUS data on 
$F_2^b$ with calculations of the 
4-flavor ZMVFN scheme, the 3-flavor scheme, and the 
BMSN prescription for the GMVFN scheme given in Figure~\ref{fig:f2b}.} 

Also the inclusive structure function $F_2$ is sensitive to the 
choice of the heavy quark scheme, due to the significant charm contribution in 
the small $x$ region. In fact, for $F_2$ the sensitivity is much larger 
than for the heavy quark contributions $F_2^{c}$ and  $F_2^{b}$ alone due to the far higher 
accuracy of the data.
In Figure~\ref{fig:f2incl} we compare the errors in $F_2$ measured by 
the H1 collaboration~\cite{h1incl} with the difference between  $F_2^{h, \rm exact} - 
F_2^{h, \rm BMSN}$. For  
$b$-quark production the scheme variation effect, 
{which is calculated in the same way as in Figure~\ref{fig:f2b},}
is negligible as compared to 
the accuracy of the data in the entire phase space.  
For the $c$-quark contributions, maximal sensitivity to the scheme choice 
appears at largest values of $Q^2$ at $x\sim 0.001$, similarly to the case of 
the data for $F_2^c$ given in Figure~\ref{fig:f2c}. The effect is 
localized in phase space and appears to be at the margin of the 
statistical resolution. Therefore the impact of the scheme variation on the 
data analysis turns out to be rather mild. To check it in a 
{more} quantitative way 
we compare the QCD-analysis of the  
inclusive DIS data performed in the 3-flavor scheme with the one in the 
BMSN prescription of the GMVFN scheme. Details and results of these analyses 
are described in the following Section.   

%%%%%%%%%%%%%%%%%%%%%%%%%%%%%%%%%%%%%%%%%%%%%%%%%%%%%%%%%%%%%%%%%%%%%%%%%%%%%%%%
\section{Impact of the Scheme Choice on the PDFs}
\label{sec4}
%%%%%%%%%%%%%%%%%%%%%%%%%%%%%%%%%%%%%%%%%%%%%%%%%%%%%%%%%%%%%%%%%%%%%%%%%%%%%%%%

\vspace*{1mm}\noindent 
We determine the PDFs from the inclusive DIS world data obtained at 
the HERA collider and in the fixed-target experiments~\cite{h1incl,Whitlow:1992uw}. These 
data are supplemented by the fixed-target Drell-Yan data~\cite{e605} and the di-muon data 
from (anti)neutrino-nucleon DIS~\cite{Goncharov:2001qe}, which allow the flavor separation 
of the sea quark distributions. 
\setcounter{table}{0} \begin{table}[h]%{angle=90} 
\begin{center} \begin{tabular}{|c|r|r|r|r|r|r|} \hline \multicolumn{1}{|c|}{ } & 
\multicolumn{1}{c|}{$a$} & \multicolumn{1}{c|}{$b$} & \multicolumn{1}{c|}{$\gamma_1$} & 
\multicolumn{1}{c|}{$\gamma_2$} & \multicolumn{1}{c|}{$A$} \\ \hline $u_v$ & $0.662 \pm 
0.034$ & $3.574 \pm 0.078$ & $-0.590 \pm 0.027$ & $-0.71 \pm 0.17$ & \\ $d_v$ & $1.06 \pm 
0.12$ & $6.42 \pm 0.41$ & $4.4 \pm 1.0$ & $-7.0 \pm 1.3$ & \\ $u_s$ & $-0.216 \pm 0.011$ & 
$6.83 \pm 0.24$ & $0.64 \pm 0.29$ & & $0.1408 \pm 0.0079$ \\ $\Delta$ & $0.7$ & $11.7 \pm 
1.9$ & $-3.5 \pm 2.1$ & & $0.256 \pm 0.082$ \\ $s$ & $-0.253 \pm 0.058$ & $7.61 \pm 0.65$ & 
& & $0.080 \pm 0.016$ \\ $G$ & $-0.214 \pm 0.013$ & $7.95 \pm 0.15$ & $0.65 \pm 0.92$ & & \\ 
\hline \end{tabular} \caption[]{\small The parameters of the PDFs and their $1\sigma$ errors 
in the 3-flavor scheme.} \label{t1} \end{center} \end{table} 
Details of the data selection, 
the corrections applied to the data, and statistical procedures used in the analysis can be 
found in Refs.~\cite{alekhin,alekhin1}. The analysis is performed taking into account the 
NNLO corrections for the light flavor contributions. 

For the neutral-current $c$-quark 
contributions to the structure function $F_2$ two variants are compared, both up to the 
level of the $O(a_s^2)$ corrections.~\footnote{The effects of the $O(a_s^3)$ 
corrections calculated recently in Ref.~\cite{Bierenbaum:2009mv} will be 
studied in a forthcoming paper.} In one case we employ 
$F_2^{c,\rm exact}(N_f=3)$ of 
Eq.~(\ref{eqFimain}), calculated for three light quark flavors choosing the factorization 
scale of $\mu^2=Q^2+4m_c^2$. 
This is compared to the BMSN prescription of the GMVFN 
scheme $F_2^{c, \rm BMSN}(N_f=4)$ given by Eq.~(\ref{eqn:BMSN}) with the factorization scale 
$\mu^2=Q^2$.
However, in the kinematical region of the data 
this variation of scale yields no difference in the fit results. Our 
fit is based on the {\it reduced cross sections} rather than the DIS structure functions. 
Therefore we also have to consider the longitudinal structure function $F_L$. Since the data are much 
less sensitive to $F_L$ than to $F_2$ the scheme choice is unimportant for the former and in 
both variants of the fit it is calculated in the $N_f=3$ FFN scheme, 
Eq.~(\ref{eqFimain}). Likewise this is the case for the $b$-quark contribution, where the 
scheme choice is also unimportant, as one can see in the comparisons of Section~\ref{sec3}. 
The $N_f=3$ FFN scheme is used both for $F_2^b$ and $F_L^b$. The charged-current 
$c$-quark contribution to the structure functions, which are related to the di-muon 
(anti)neutrino-nucleon DIS data used in the fit, are calculated in the $N_f = 3$ 
FFN scheme at NLO~\cite{Gottschalk:1980rv}.
For the Drell-Yan cross sections, we include the NNLO QCD corrections~\cite{dyrap}. 
In this case the 5-flavor PDFs defined in Eqs.~(\ref{eqn:hqpdf}--\ref{HPDF4})
are used in order to take into account the $c$- and 
$b$-quark contributions. Note, however, that at the typical fixed-target 
energies the impact of heavy quarks is marginal and the 3-flavor scheme
provides a sufficiently good description.

%
%\newpage
\begin{figure*}%[h]
\vspace*{-3cm} \hspace*{3cm}
\includegraphics[angle=0, height=23.0cm]{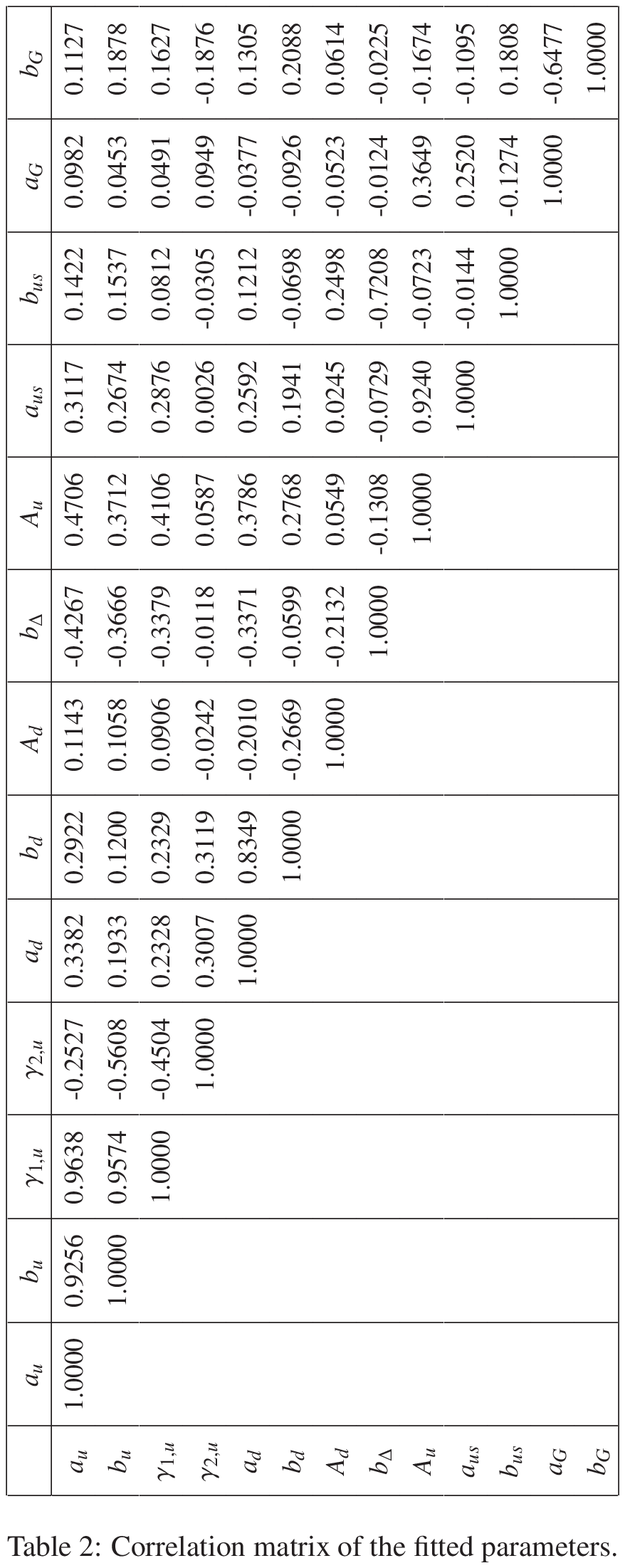}
\end{figure*}
%%%%%%%%%%%%%%%%%%%---------------------------------------------------------------
%
%\newpage
\begin{figure*}%[h]
\vspace*{-3cm} \hspace*{3cm}
\includegraphics[angle=0, height=23.0cm]{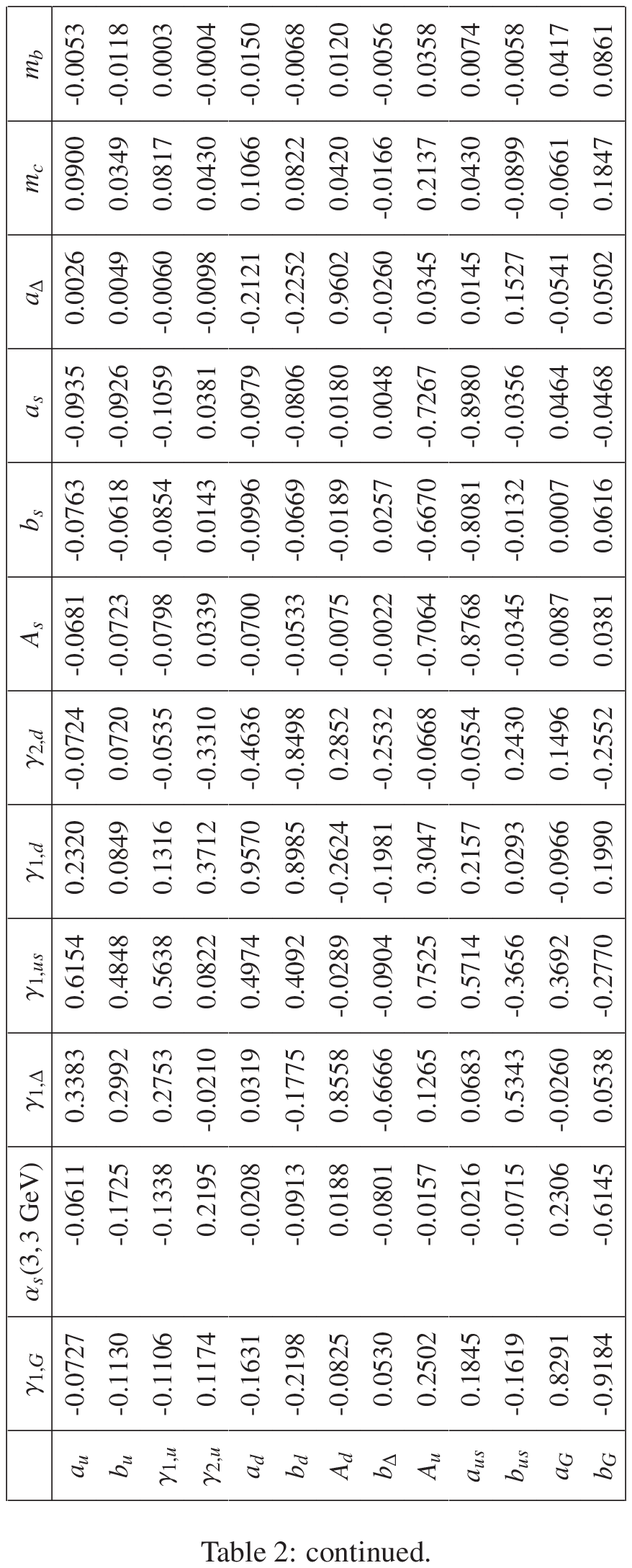}
\end{figure*}
%%%%%%%%%%%%%%%%%%%---------------------------------------------------------------
%
%\newpage
\begin{figure*}%[h]
\vspace*{-3cm} \hspace*{3.3cm}
\includegraphics[angle=0, height=23.0cm]{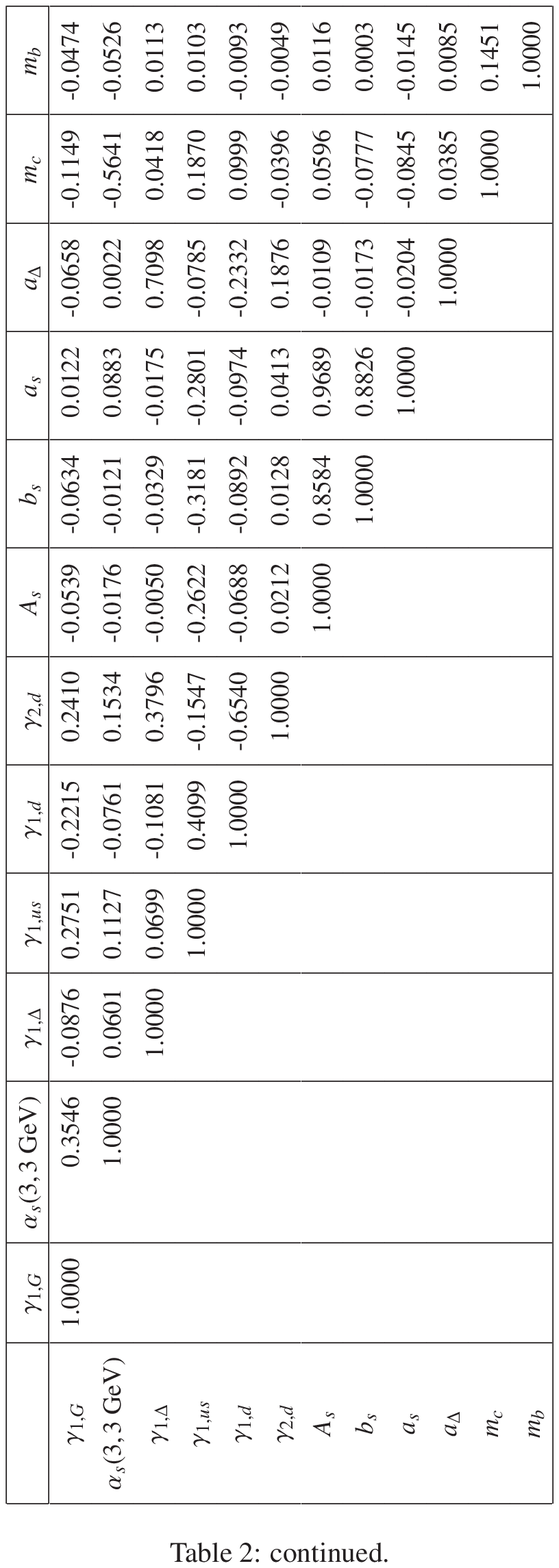}
\end{figure*}
%%%%%%%%%%%%%%%%%%%---------------------------------------------------------------
\setcounter{table}{2}
%\noindent
%
The proton PDFs are parameterized at the scale  $Q_0^2=9~{\rm GeV}^2$ 
in the 3-flavor scheme. At the starting scale
the following functions are used for the valence quark, gluon, and sea-quark 
distributions~:
%-------------------------------------------------------------------------------
\begin{eqnarray}
xq_{V}(x,Q_0^2)&=&\frac{ 2 \delta_{q{u}}+\delta_{q{d}}}
{N^{V}_q}x^{a_{q}}(1-x)^{b_q}x^{P_{q,V}(x) },
~~~P_{q,{V}} = \gamma_{1,q} x + \gamma_{2,q} x^2,
~~~q = u,d~,
\label{eqn:pdf1}\\
xG(x,Q_0^2)&=&A_{G}x^{a_{G}}(1-x)^{b_{G}}
x^{P_{G}(x) },~~~P_{G} = \gamma_{1,{G}} x~,
\label{eqn:pdf5}\\
xu_{S}(x,Q_0^2)&=&x\bar{u}_{S}(x,Q_0^2)
=A_u x^{a_{u{s}}}(1-x)^{b_{u{s}}}
x^{P_{u,s}(x) },~~~P_{u,{s}} = \gamma_{1,{u{s}}} x~,
\label{eqn:pdf2}\\
x\Delta(x,Q_0^2)&=&xd_{S}(x,Q_0^2)-xu_{S}(x,Q_0^2)
=A_\Delta x^{a_{\Delta}}(1-x)^{b_{\Delta}}
x^{P_{\Delta}(x) },~~~P_{\Delta} = \gamma_{1,{\Delta}} x~.
\label{eqn:pdf3}
\end{eqnarray}
%-------------------------------------------------------------------------------
The strange quark distribution is taken in the charge-symmetric form 
\begin{equation}
xs(x,Q_0^2)=x\bar{s}(x,Q_0^2)=A_{s} x^{a_{s}}(1-x)^{b_{s}}, 
\label{eqn:pdf4}
\end{equation}
in agreement with the results of Ref.~\cite{alekhin1}. 
The polynomials $P(x)$ used in Eqs.~(\ref{eqn:pdf1}--\ref{eqn:pdf3})
provide sufficient flexibility of 
the PDF-parameterization with respect to the analyzed data 
and no additional terms are required to 
improve the fit quality. The PDF-parameters determined from the fit performed 
in the 3-flavor scheme are given in Table~\ref{t1}. 
Because of the lack of the neutron-target data in the region of small values of  
$x$, the low-$x$ exponent $a_{\Delta}$ cannot be defined from the fit  
and we fix it to 0.7 to choose an ansatz, in agreement with the values obtained for 
the 
low-$x$ exponents of the valence quark distributions and phenomenological 
estimates, 
cf. e.g.~\cite{Ermolaev:2000sg}.
However, once we have fixed $a_{\Delta}$ the uncertainty in the sea quark 
distributions at small $x$ is underestimated. 
We therefore choose an uncertainty  $\delta a_{\Delta} = 0.3$, and, 
in order to account for its impact on the other PDF-parameters, we calculate 
the errors in the latter with the value of $a_{\Delta}$ released, but 
with an additional pseudo-measurement of $a_{\Delta}=0.7 \pm 0.3$
added to the data set. 
\noindent
In our fit the heavy quark masses are fixed at 
$m_c=1.5~{\rm GeV}$ and $m_b=4.5~{\rm GeV}$ and
the same approach is employed to take into account 
possible variations of $m_c$ and $m_b$
in the ranges of $\pm0.1~{\rm GeV}$ and $\pm0.5~{\rm GeV}$, respectively.
Note that the normalization parameters 
for the valence quarks and gluons are defined from other PDF-parameters 
applying both fermion number- and momentum conservation. In the global fit we 
obtain 
%-------------------------------------------------------------------------------
\begin{equation}
\frac{\chi^2}{\rm NDP} = \frac{3038}{2716} = 1.1
\end{equation}
%-------------------------------------------------------------------------------
for the parameter values listed in Table~\ref{t1}.  
\begin{figure}[htb]
  \begin{center}
    \includegraphics[width=16.5cm]{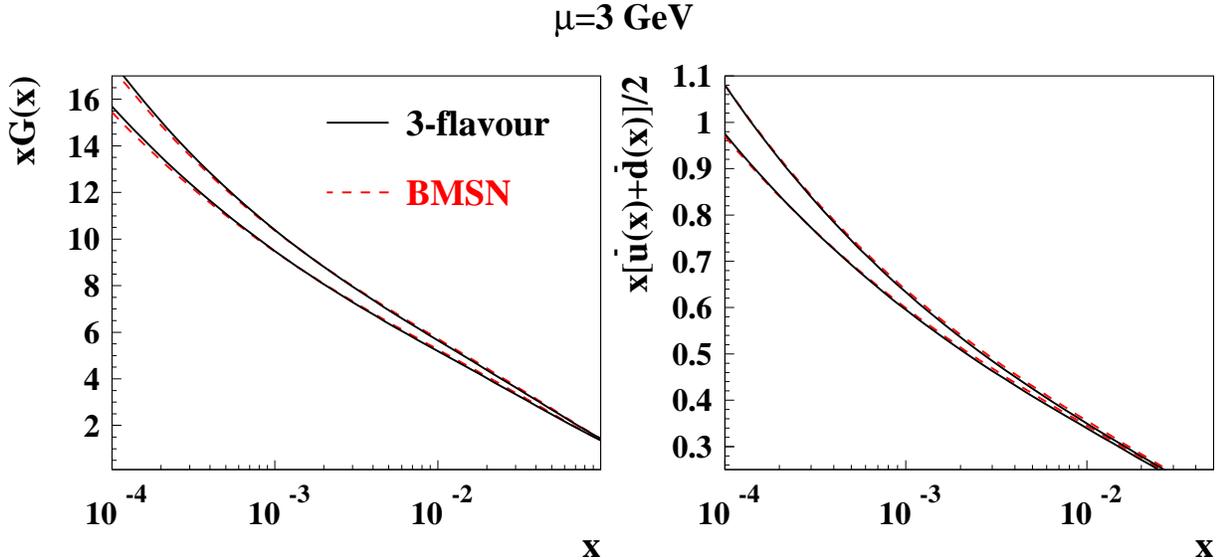}
    \vspace*{-5mm}
    \caption[]{ \small The $1\sigma$ error band for the gluon 
      (left panel) and sea (right panel) 
      distributions obtained in two variants of the fit. Solid lines: 3-flavor 
      scheme, dashed lines : GMVFN scheme in the BMSN prescription.
      \label{fig:pdfcomp}
    }
  \end{center}
\end{figure}

In the fit 25 
parameters are determined. 
The covariance 
matrix elements for these parameters are given in Table~2. %\ref{tab:cov}.
%------------------------------------------------------------------------------
\begin{figure}[htb]
  \begin{center}
    \includegraphics[width=11.0cm]{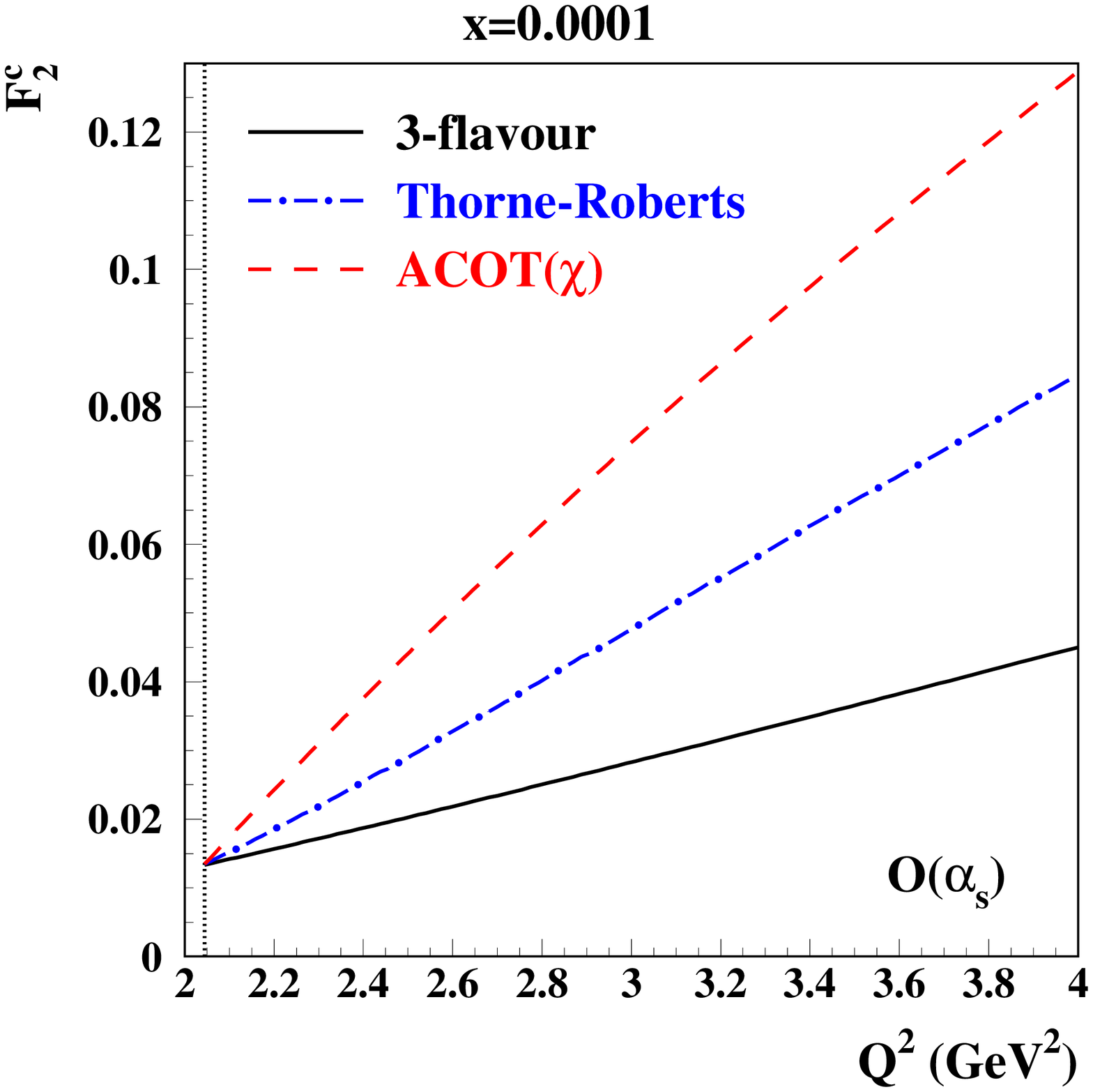}
    \vspace*{-5mm}
    \caption[]{ \small
     Matching of $F_2^{c, \rm TR}(x,Q^2)$
     (dash-dotted line) and $F_2^{c,{\rm ACOT}(\chi)}(x,Q^2)$ (dashed line) 
     with $F_2^{c,\rm exact}(N_f=3,x,Q^2)$ (solid line) at small $Q^2$
     at $O(\alpha_s)$. The MRST2001 PDFs of Ref.\cite{Martin:2002aw} are 
     used.  The vertical line denotes the  
     position of the charm-quark mass $m_c=1.43~{\rm GeV}$.
      \label{fig:gmvfn}
    }
  \end{center}
\end{figure}
%------------------------------------------------------------------------------
\noindent
The parameter errors quoted
are due to the propagation of the statistical and systematic errors in the data.
The
error correlations are taken into account if available, which is
the case for most of the data sets considered.

The gluon and flavor singlet distributions 
obtained in case of the BMSN prescription 
are compared to those referring to the 3-flavor scheme 
in Figure~\ref{fig:pdfcomp}.
The difference between the two variants is quite small and 
situated well within 
the PDF-uncertainties. For the non-singlet PDFs it is even smaller, since the 
heavy quark contribution is negligible at $x\gtrsim 0.1$, 
cf. Ref.~\cite{Blumlein:2006be}. 
For the BMSN variant of the fit a value of $\chi^2/{\rm NDP} = 3036/2716$ is obtained, 
very 
close to 
the one for the fit in the 3-flavor scheme. 
This is in line with the comparisons given in Section~\ref{sec3}, which 
show that in the case of a smooth matching of the 3-flavor 
and VFN scheme at small values of  
$Q^2$, there is little room for a difference between them in the region of the
present experiments. 

\begin{figure}[t]
  \begin{center}
    \includegraphics[width=16.5cm]{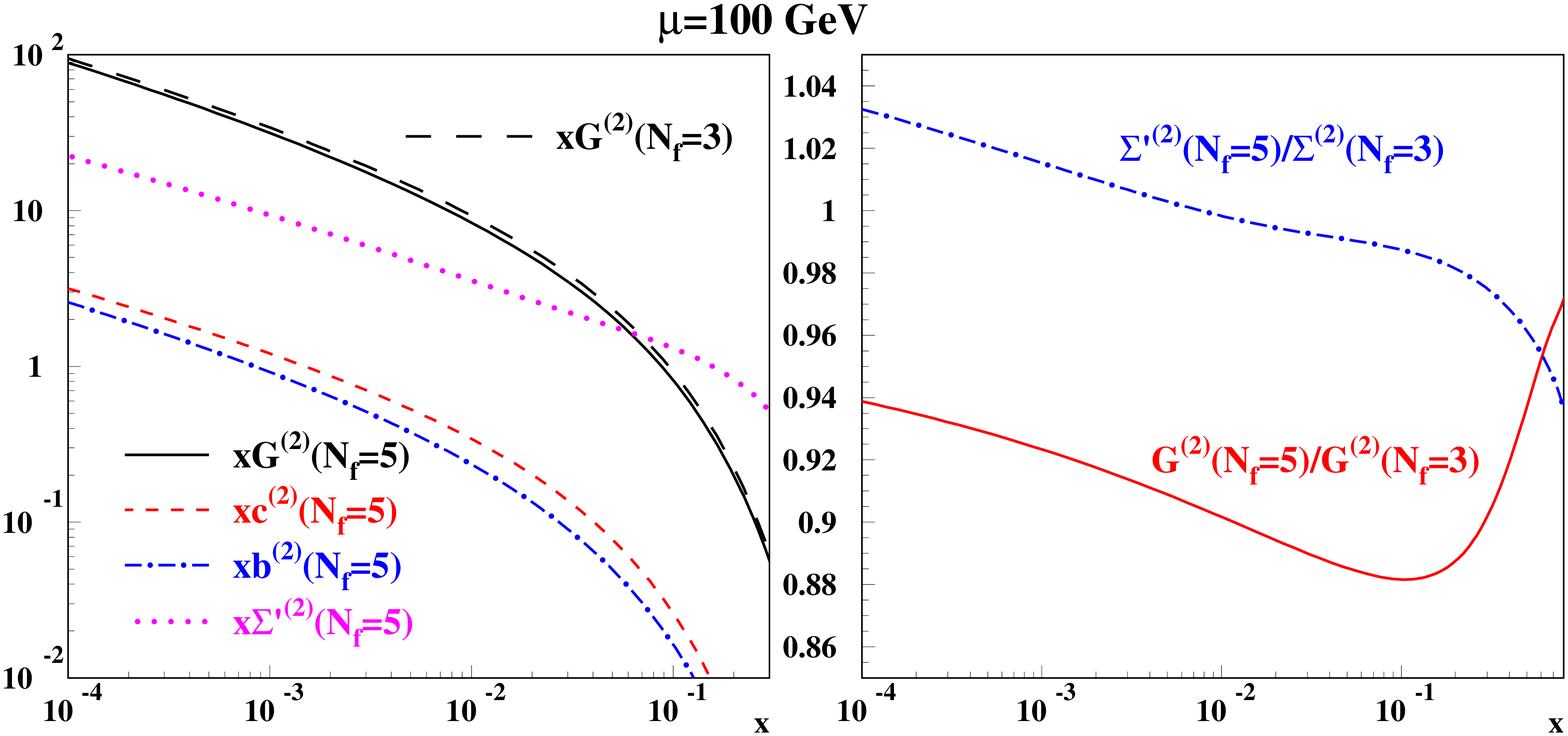}
    \vspace*{-5mm}
    \caption[]{ \small Left panel: The 5-flavor PDFs 
    at the scale of $\mu^2=10^4~{\rm GeV^2}$ and the 3-flavor  
    gluon distributoin given for comparison.
    Right panel: The ratio of the 
    5-flavor to 3-flavor gluon (solid line) and singlet (dash-dotted line) 
    distributions at the same scale.   
       \label{fig:vfnpdfs}
    }
  \end{center}
\end{figure}

\setcounter{table}{2}
\begin{table}[h]
\renewcommand{\arraystretch}{1.3}
\begin{center}
\begin{tabular}{|l|r|l|}
\hline
\multicolumn{1}{|c|}{ } &
\multicolumn{1}{c|}{$\alpha_s({\rm M_Z^2})$} &
\multicolumn{1}{c|}{  } \\
\hline
ABKM           & $0.1135 \pm 0.0014$ & {\rm heavy~quarks:~FFN~$N_f=3$}             
\\
ABKM           & $0.1129 \pm 0.0014$ & {\rm heavy~quarks:~BMSN approach}             
\\
BBG \cite{Blumlein:2006be} & $
0.1134 {\footnotesize{\begin{array}{c} +0.0019 \\
-0.0021 \end{array}}}$
& {\rm valence~analysis, NNLO}             \\
AMP06 \cite{alekhin} & $0.1128 \pm 0.0015$ &
\\
JR \cite{JR}& $0.1124 \pm 0.0020$ & {\rm dynamical~approach}             
\\
MSTW \cite{Martin:2009bu} & $0.1171\pm 0.0014$ &         \\     
\hline
BBG \cite{Blumlein:2006be} & $
0.1141 {\footnotesize{\begin{array}{c} +0.0020 \\
-0.0022 \end{array}}}$
& {\rm valence~analysis, N$^3$LO}             \\
\hline
\end{tabular}
\caption[]{\small Comparison of different measurements of $\alpha_s(\rm M_Z^2)$ 
at NNLO and higher order.}
\label{Tab:alpha}
\end{center}
\renewcommand{\arraystretch}{1}
\end{table}
%
%--------------------------------------------------------------------------------
\begin{figure}[h]
  \begin{center}
    \includegraphics[width=10cm]{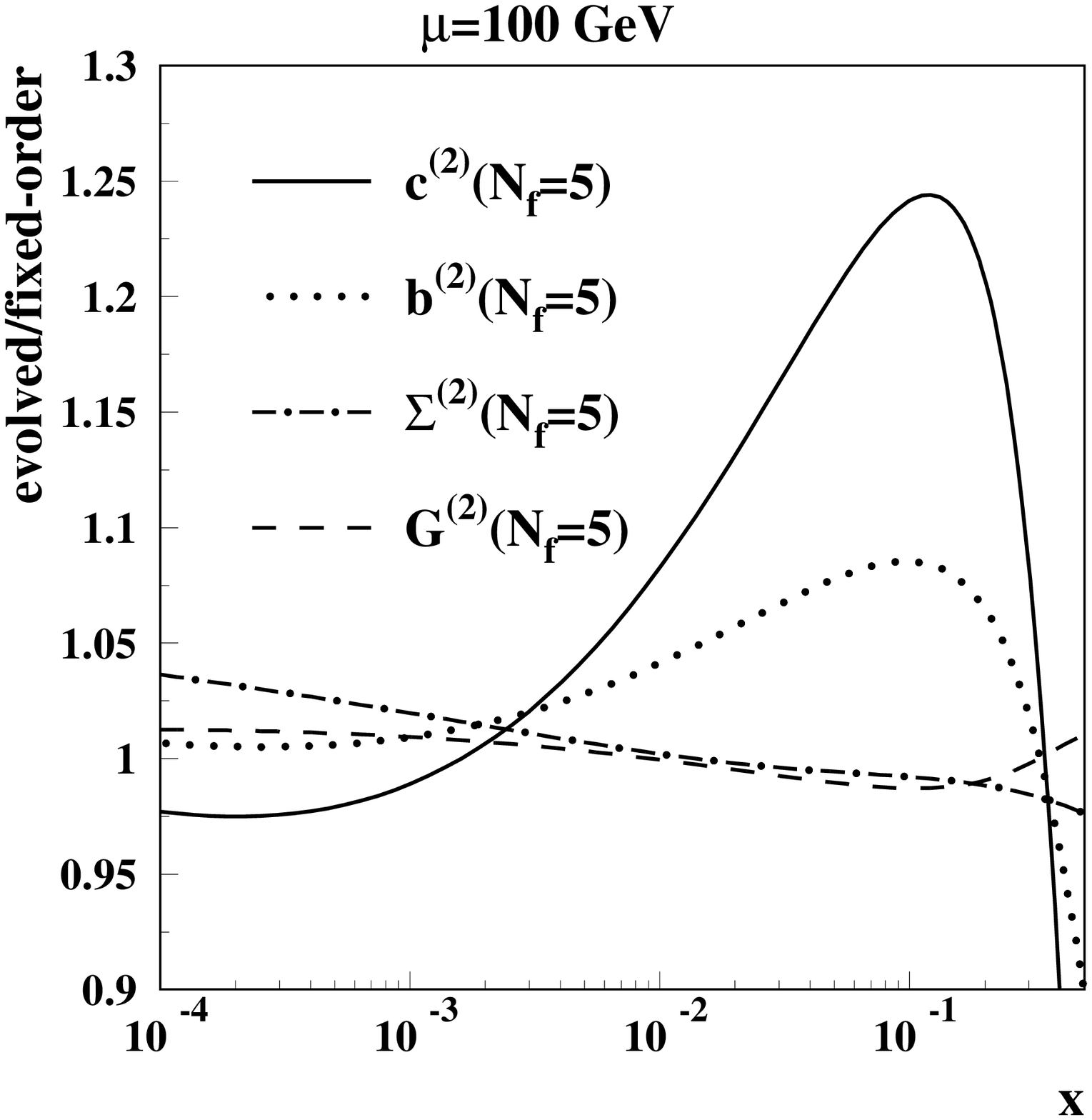}
    \vspace*{-5mm}
    \caption[]{ \small Ratio of the evolved NNLO 5-flavor distributions to 
              the distributions being obtained applying 
              the fixed-order matching conditions of 
              Eqs.~(\ref{eqn:hqpdf}--\ref{HPDF4}). 
       \label{fig:evolfull}
    }
  \end{center}
\end{figure}
%--------------------------------------------------------------------------------

The difference between the fits performed using the TR-prescription for the
GMVFN scheme and in the 3-flavor scheme is also not dramatic, as one can 
see in the ZEUS NLO PDF fit, Ref.~\cite{CooperSarkar:2007ny}. However, 
it is somewhat larger than in the case of the BMSN prescription. In 
particular, 
this happens since the TR-prescription does not provide a smooth matching 
with the 3-flavor scheme at low values of $Q^2$ and at small values of $x$. 
By construction, the TR-prescription provides a smooth transition for 
the {\it gluon-initiated} contribution only. However, at small values of $Q^2$ 
the gluon distribution has a valence-like form and falls at small $x$. 
As a result, the quark-singlet contribution to the slope  
$\partial F_2^c/\partial \ln(Q^2)$ is non-negligible at small values 
of 
$x$, which leads to a kink at the matching point $Q^2 =m_h^2$ in  
$F_2^h$ using the TR-prescription, see Figure~\ref{fig:gmvfn}.   
This is an artefact of the description leading to an overestimation of 
the heavy quark contribution and,  correspondingly, an underestimation of 
the fitted light quark PDFs at small values of $x$.
In the ACOT($\chi$)-prescription the smoothness is not required by definition 
and the kink in $F_2^h$ is even bigger than for the case of TR-prescription. 
In the most recent version of the ACOT-prescription this problem is 
addressed~\cite{Nadolsky:2009ge} and  should yield a result 
closer to the 3-flavor scheme than the ACOT($\chi$)-prescription.   

Summarizing the comparisons of Sections~\ref{sec3} and \ref{sec4} 
we conclude that, once the $O(\alpha_s^2)$ corrections to heavy quark 
electro-production are taken into account and a smooth matching with the 3-flavor 
scheme at small $Q^2$ is provided, the GMVFN scheme should agree to the 3-flavor 
scheme for the kinematics explored by experiments so far. 
Furthermore, it is expected that the NNLO corrections to the heavy quark
structure functions of Ref.~\cite{Bierenbaum:2009mv} lead to an even better agreement.

For the strong coupling constant at NNLO in QCD the values 
\begin{eqnarray}
\label{eq:alphas-FFN}
\alpha_s^{\overline{\rm MS}}(N_f=5, M_Z^2) &=& 0.1135 
\pm 0.0014~{\rm (exp)}
\hspace{5mm}~{{\rm FFN~scheme,}~~~N_f = 3}
\\
\label{eq:alphas-VFN}
\alpha_s^{\overline{\rm MS}}(N_f=5, M_Z^2) &=& 0.1129 
\pm 0.0014~{\rm (exp)}
\hspace{5mm}~{\rm BMSN scheme}
\end{eqnarray}
are obtained.
The small difference between these two values lies well within the experimental
uncertainty. In Table~\ref{Tab:alpha} we compare these values to other recent NNLO determinations 
of the strong coupling constant. 
Our results agree very well with those of Refs.~\cite{Blumlein:2006be,JR}. 
Note that the data sets used in the 
non-singlet fit of Ref.~\cite{Blumlein:2006be} are rather 
different from those used in the present analysis. 
The value of $\alpha_s(M_Z^2)$ given in
Ref.~\cite{Martin:2009bu} is by $2.7 \sigma$ larger. As is well-known from the 
non-singlet 
data analysis \cite{Blumlein:2006be}, a somewhat higher value of $\alpha_s(M_Z^2)$ is 
obtained at N$^3$LO, cf. also Ref.~\cite{Vermaseren:2005qc} for an estimate. The 
difference of these determinations at NNLO and N$^3$LO is half of 
the experimental error found in the present analysis. Eqs.~(\ref{eq:alphas-FFN}) and (\ref{eq:alphas-VFN}) determine $\alpha_s(M_Z^2)$ 
at an accuracy of $\approx 1.5\%$. 

%%%%%%%%%%%%%%%%%%%%%%%%%%%%%%%%%%%%%%%%%%%%%%%%%%%%%%%%%%%%%%%%%%%%%%%%%%%%%%%%%%
\section{Applications to Collider Phenomenology}
\label{sec5}
%%%%%%%%%%%%%%%%%%%%%%%%%%%%%%%%%%%%%%%%%%%%%%%%%%%%%%%%%%%%%%%%%%%%%%%%%%%%%%%%%%

\vspace*{1mm}\noindent 
In this Section, we investigate the implications of the PDFs obtained
in the present NNLO analysis for collider phenomenology.
To that end we focus on important (semi)-inclusive scattering cross sections
at hadron colliders, such as the Drell-Yan process for $W^{\pm}$- and $Z$-boson production,
the pair-production of top-quarks and (Standard Model) Higgs-boson production.
The corresponding cross sections in, say, proton-proton scattering can be
written as
\begin{eqnarray}
\label{eq:QCDfactorization}
  \displaystyle
  \sigma_{pp \to X}(s) =
  \sum\limits_{ij}\,
  \int
  dx_1\, dx_2\,
  f_{i}(x_1,\mu^2) \,
  f_{j}(x_2,\mu^2) \,
  \hat{\sigma}_{ij \to X} \left(x_1, x_2, s,\alpha_s(\mu^2),\mu^2 \right)\,
  \, ,
\end{eqnarray}
where $X$ is the final state under consideration and $s$ the c.m.s. energy. 
The PDFs are collectively denoted by $f_i$, $f_j$ and the sum runs over all partons.
At hadron colliders the convolution of $f_i$ and $f_j$ parameterizes the
so-called parton luminosity $L_{ij}$.
In the following, we will employ our PDFs and present numbers
for $p \bar p$-collisions at Tevatron with $\sqrt{s} = 1.96$~TeV
and for $pp$-collisions at LHC at energies $\sqrt{s} = 7, 10$ and $14$~TeV.
To that end, we have to rely on the perturbative QCD evolution of the light
and heavy PDFs to Tevatron and LHC scales, which puts us also in the position
to compare to other global PDF analyses. In the comparison we will consider
the impact of the error of the PDFs at the level $1\sigma_P$ (the index $P$ 
denoting PDFs), which results from the 
experimental errors in the PDF analysis, including full error correlation, 
see Section~\ref{sec4}. 
We will not consider theory errors implied by varying the
factorization and renormalization scales. 
At the level of NNLO these amount typically to a few per cent only and,
moreover, are largely independent of the PDFs.
Also the anticipated statistical and systematic errors
in the measurements at Tevatron and the corresponding resolutions, which can be achieved at the 
LHC, are not considered. 

%---------------------------------------------------------------------------------
\subsection{Evolution of Light and Heavy PDFs}
\label{sec51}
%---------------------------------------------------------------------------------

\vspace*{1mm}\noindent
The typical energy scales for hard scattering processes 
at high-energy hadron colliders are often much larger than the $c$-quark mass, 
and even than the $b$-quark mass. In this case the $(4-)5-$flavor scheme 
is the relevant choice, if power corrections and non-factorizing 
contributions can be safely neglected. Moreover, very often this is the only 
approach feasible, since the cross sections of the partonic sub-processes are 
only available in the approximation of massless initial-state partons. 
The 3-flavor PDFs obtained from the fit in Section~\ref{sec4} can be used 
to generate the 4-flavor distributions using the matching conditions in 
Eqs.~(\ref{eqn:hqpdf}--\ref{HPDF4}). As we show in 
Figure~\ref{fig:evol}, at $O(\alpha^2_s)$ and low scales, the PDFs 
computed in this way are very similar to the evolved ones, provided the matched PDFs 
are taken as boundary conditions in the evolution. 
At large scales the difference between these two cases is non-negligible,
contrary to the case of heavy quark DIS electro-production. The large-log 
resummation effects can be important in some range of the phase space at 
hadron colliders. In view of these aspects we obtain the 4-flavor PDFs from 
the NNLO evolution with the boundary scale  $m_c^2$ 
and the boundary conditions given in Eqs.~(\ref{eqn:hqpdf}--\ref{HPDF4}). The 
5--flavor PDFs are obtained from the evolved 4-flavor distributions using 
analogous boundary conditions at the scale  $m_b^2$. Since at $m_b^2$ 
the mass effects of the charm-quark are not negligible due to 
$m_b^2/m_c^2 \sim 
10$, this approach is {\it some approximation}, the validity of which has to be 
tested for the corresponding processes. The problem of the heavy quark mass 
scale separation cannot be resolved within the concept of a VFN scheme and
implies an unavoidable theoretical uncertainty related to the use of VFN 
PDFs.%
%--------------------------------------------------------------------------------
\begin{figure}[htb]
  \begin{center}
    \includegraphics[width=16.5cm]{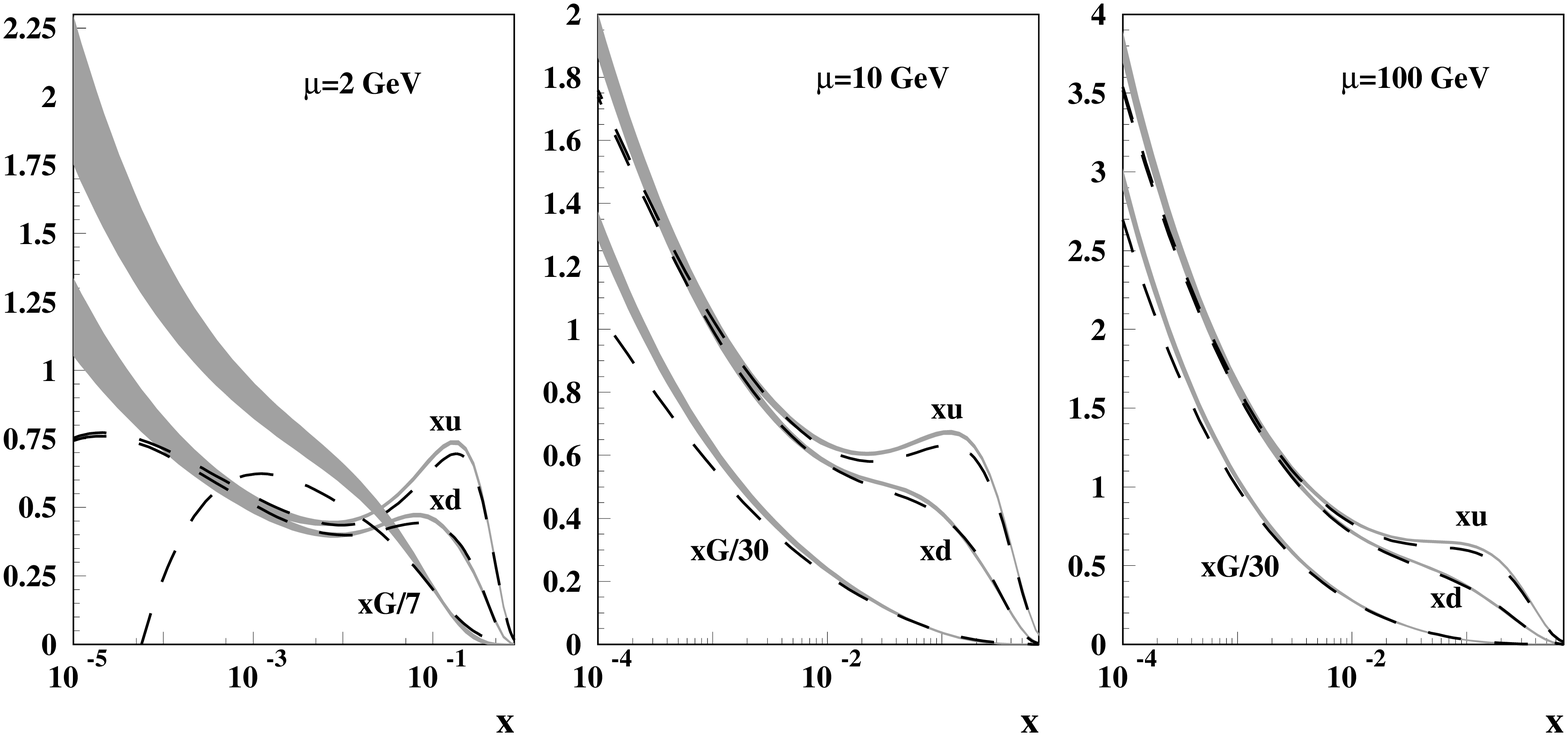}
    \vspace*{-5mm}
    \caption[]{ \small The $1\sigma$ error bands (shaded area) for 
    our NNLO 4-flavor (left panel) 
    and 5-flavor (central and right panels) $u$-, $d$-quark, and gluon 
    distributions in comparison to the corresponding MSTW2008 NNLO 
    distributions \cite{Martin:2009iq} (dashed lines).    
       \label{fig:comp}
    }
  \end{center}
\end{figure}
%--------------------------------------------------------------------------------
\begin{figure}[htb]
  \begin{center}
    \includegraphics[width=16.5cm]{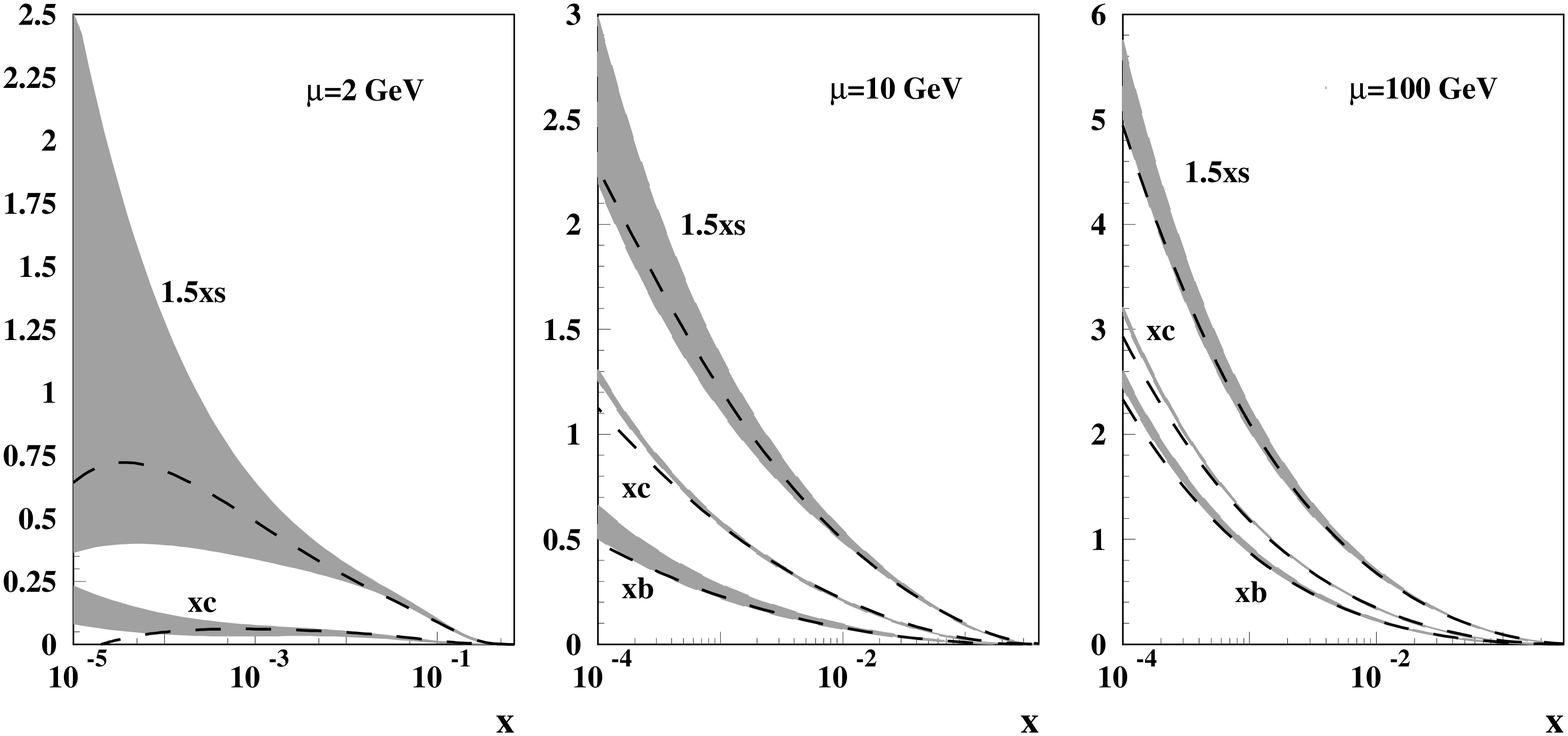}
    \vspace*{-5mm}
    \caption[]{ \small 
    The $1\sigma$ error bands (shaded area) for
    our NNLO 4-flavor (left panel)
    and 5-flavor (central and right panels) $s$-, $c$-, and $b$-quark
    distributions in comparison to the corresponding MSTW2008 NNLO
    distributions \cite{Martin:2009iq} (dashed lines).
       \label{fig:comp1}
    }
  \end{center}
\end{figure}
%--------------------------------------------------------------------------------
%--------------------------------------------------------------------------------
\begin{figure}[htb]
  \begin{center}
    \includegraphics[width=16.5cm]{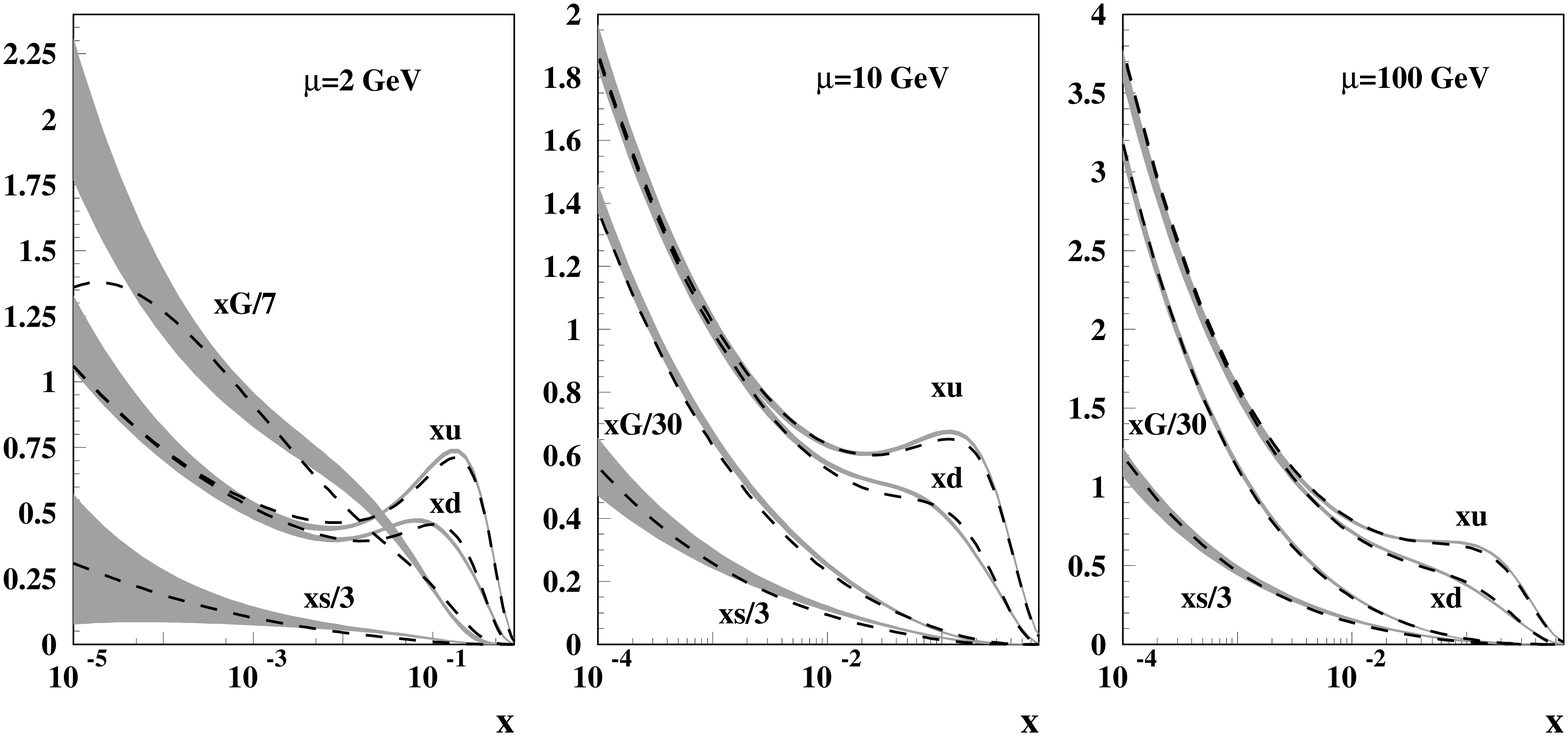}
    \vspace*{-5mm}
    \caption[]{ \small 
       \label{fig:comp2}
The $1\sigma$ error bands (shaded area) for
    our NNLO 3-flavor $u$-, $d$-, $s$-quark and gluon 
    distributions in comparison to the corresponding JR NNLO
    distributions \cite{JR} (dashed lines).
    }
  \end{center}
\end{figure}
%--------------------------------------------------------------------------------
In general, the heavy quark PDFs rise with the 
scale $\mu^2$, while the $(N_f+1)$-light PDFs decrease correspondingly
with respect to $N_f$-light PDFs.
At the scale of $10^4~{\rm GeV^2}$ the 5-flavor gluons lose some 7\% of 
momentum as compared to the 3-flavor ones. This momentum is transferred 
to the $c$- and $b$-quark distributions, see Figure~\ref{fig:vfnpdfs}. 
The difference between the 3-flavor singlet distribution $\Sigma(N_f=3)$
and the 5-flavor distribution
\begin{eqnarray}
\Sigma' \left(N_f=5\right) =
\sum_{k=1}^{3}\left[ q_k\left(N_f=5\right)+\bar{q}_k\left(N_f=5\right)\right] 
\end{eqnarray}
is smaller than that for the gluons since in  the quark-case the corresponding OMEs 
appear only at $O(\alpha^2_s)$. 
At small values of $x$ the 5-flavor 
light quark and gluon distributions receive an additional enhancement as 
compared to the 3-flavor distributions due to evolution, 
see~Figure~\ref{fig:evolfull}. 
{This difference can be considered as an estimate of the theoretical 
uncertainty in the 5-flavor PDFs due to the higher order corrections.}
For the $c$- and $b$-quark distributions 
at $x \sim 0.1$ the effect of the evolution is much larger. However, 
due to the smallness of the heavy quark PDFs in this region its absolute   
magnitude is insignificant for most practical purposes. 
%---------------------------------------------------------------------------------
\subsection{Comparison with Other NNLO Analyses}
\label{sec52}
%---------------------------------------------------------------------------------

\vspace*{1mm}\noindent
In Figures~\ref{fig:comp} and \ref{fig:comp1} we compare the NNLO PDFs 
obtained in the present analysis to the PDFs by Martin-Stirling-Thorne-Watt of 2008
(MSTW 2008), \cite{Martin:2009iq}. 
At the scales of $\mu^2=100~{\rm GeV^2}$ and $\mu^2=10^4~{\rm GeV}$,
we compare the 5-flavor PDFs and at the scale of $\mu^2=4~{\rm GeV}^2$,
the 4-flavor PDFs since for the MSTW2008 set the number of flavors is four
at $m_c< \mu<m_b$ and 5 at $\mu>m_b$~\footnote{If the scale 
is not much larger than $m_c^2$ the choice of 3-flavor PDFs is 
most relevant, cf.~Sections~\ref{sec2},\ref{sec3}.}.
At small values of $x$ our gluon distribution is larger than that of  MSTW2008. 
This difference is particularly essential at smaller scales where the NNLO 
MSTW2008 gluon distribution becomes negative at $x\sim 5 \cdot 10^{-5}$. This 
is not the case 
in our analysis. Also our sea-quark distributions 
are larger than those of MSTW2008 in the small-$x$ region. As we 
discuss in Section~\ref{sec4}, this might be partly related to the 
heavy quark contribution in the GMVFN scheme employed in 
the MSTW2008 fit. The shape of the gluon distribution at small $x$ is 
sensitive to the recent measurements of $F_L$ at small $Q^2$ by the H1 and ZEUS 
collaborations~\cite{FLexp}. 
These measurements are in agreement with our shape for the gluon and 
do slightly disfavor the MSTW2008 predictions. At large $x$ the MSTW2008 
gluon distribution is somewhat larger than ours due to the impact of 
the Tevatron jet data included in the MSTW2008 analysis. 

\begin{table}[h]
\renewcommand{\arraystretch}{1.3}
\begin{center}
{\small
\begin{tabular}{|l|c|c|c|}
\hline
\multicolumn{1}{|c|}{ } &
\multicolumn{1}{c|}{$\langle x u_v(x)\rangle$} &
\multicolumn{1}{c|}{$\langle x d_v(x)\rangle$} &
\multicolumn{1}{c|}{$\langle x [u_v-d_v](x)\rangle$} \\
\hline
ABMK            
& $0.2981 \pm  0.0025$ 
& $0.1191 \pm  0.0023$  
& $0.1790 \pm  0.0023$  
\\
BBG            
\cite{Blumlein:2006be}
& $0.2986 \pm 0.0029$ 
& $0.1239 \pm 0.0026$  
& $0.1747 \pm 0.0039$  
\\
JR \cite{JR}
& $0.2900 \pm 0.0030$ 
& $0.1250 \pm 0.0050$  
& $0.1640 \pm 0.0060$  
\\
MSTW 
\cite{Martin:2009bu}
& $0.2816 
{\footnotesize
{\begin{array}{c} +0.0051 \\
                  -0.0042 \end{array}
}}$
& $0.1171 
{\footnotesize{\begin{array}{c} +0.0027 \\
                                -0.0028 
               \end{array}}}$
& $0.1645 
{\footnotesize{\begin{array}{c} +0.0046 \\
                                -0.0034 
               \end{array}}}$
\\     
AMP06 \cite{alekhin} 
& $0.2947 \pm 0.0030$ 
& $0.1129 \pm 0.0031$  
& $0.1820 \pm 0.0056$  
\\     
\hline
BBG [N$^3$LO]           
\cite{Blumlein:2006be}
& $0.3006 \pm 0.0031$  
& $0.1252 \pm 0.0027$  
& $0.1754 \pm 0.0041$ 
\\
\hline
\end{tabular}
}
\caption[text]{\small Comparison of the 2nd moment of the valence quark distributions
at NNLO and N$^3$LO obtained in different analyses at $Q^2 = 4~{\rm GeV}^2$.~\footnotemark[5]} 
\label{Tab:mom}
\end{center}
\renewcommand{\arraystretch}{1}
\end{table}
\footnotetext[5]{We thank P.~Jimenez-Delgado and
W.J. Stirling for providing us with the moments of the JR and
MSTW08 distributions.}
\setcounter{footnote}{5}
In Figure~\ref{fig:comp2} we compare our 3-flavor PDFs 
to the results obtained by the Dortmund group 
[Jimenes-Delgado and Reya (JR)] in Ref.~\cite{JR},
for $\mu^2 = 4, 100$ and $10000~{\rm GeV}^2$. 
At $\mu^2 = 4~{\rm GeV}^2$ the gluon PDF \cite{JR} is 
somewhat smaller for $x \lsim 5 \cdot 10^{-5}$ than the gluon distribution 
determined in the present fit. This is a region in which the fit is not 
constrained by data. A very small difference is also observed for 
the $u-$ and $d-$quark distributions in the region $x \sim 0.1$. 
Otherwise, one notices very good agreement of both distributions.

In Table~\ref{Tab:mom} we summarize different values of the 2nd moment of the 
valence quark densities.~\footnote{Here and in the following we restrict the 
comparison to the
results obtained in NNLO analyses. Currently available NLO analyses 
(see in Ref.~\cite{Blumlein:2006be} and Refs.~\cite{h1incl,ZEUSNLO,CTEQ,NNW}) 
contain relatively large theory uncertainties of $\pm 0.0050$ for
$\alpha_s(M_Z^2)$, much larger than the experimental accuracy presently 
reached.} 
They are closely related to the moments which are currently 
measured in lattice simulations \cite{RENNER}. 
The values of all analyses are very 
similar, with some differences still visible. A quantity of central importance is
%-----------------------------------------------------------------------------------
\begin{eqnarray}
\langle xV(Q^2)\rangle  = \int_0^1 dx~x\left\{\left[u(x,Q^2) + \bar{u}(x,Q^2)\right]  
           -       \left[d(x,Q^2) + \bar{d}(x,Q^2)\right]\right\}~.  
\end{eqnarray}
%-----------------------------------------------------------------------------------
%--------------------------------------------------------------------------------
\begin{figure}[h]
  \begin{center}
    \includegraphics[width=14.5cm]{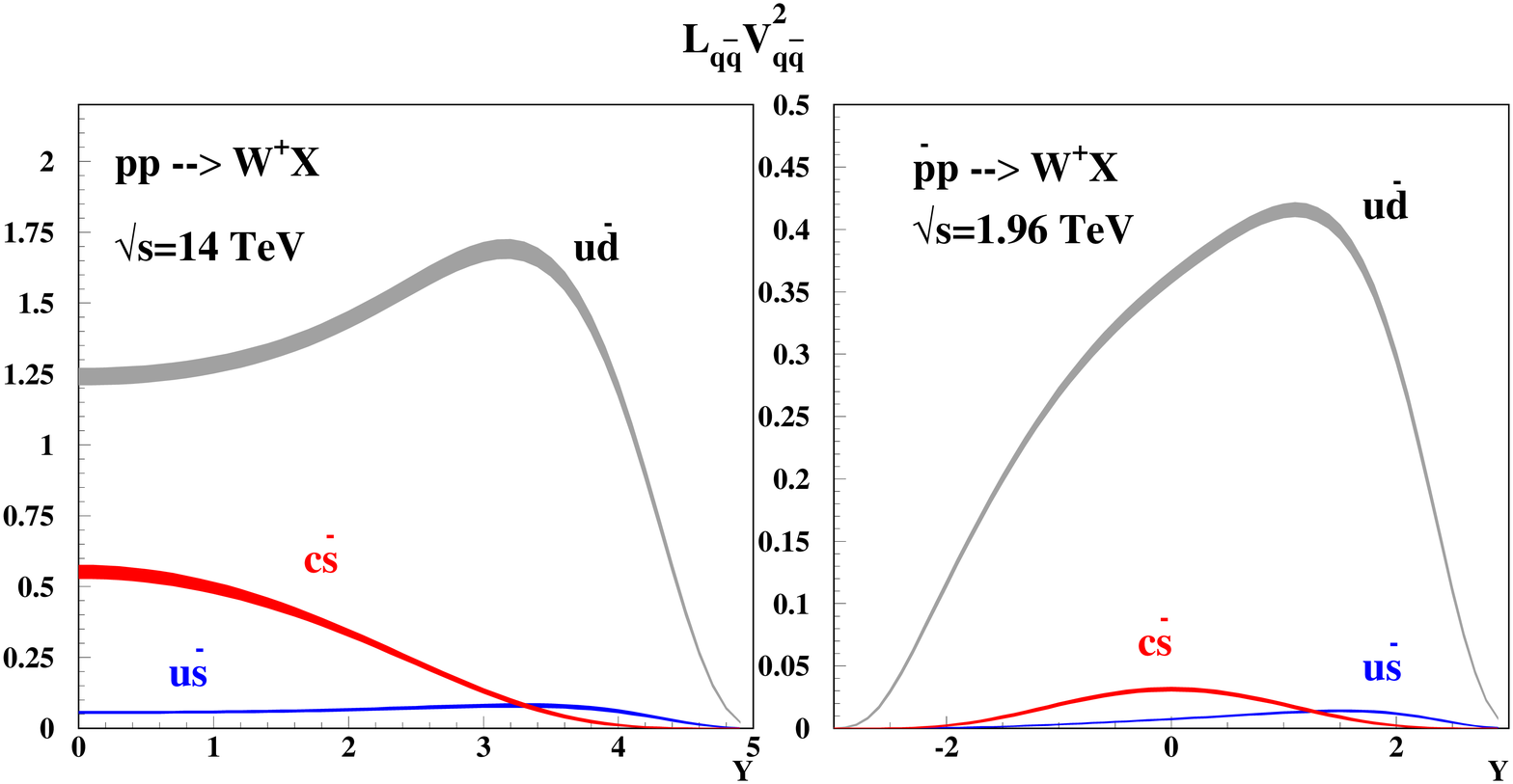}
    \vspace*{-5mm}
    \caption[]{ \small The  { 1$\sigma$ band for the} 
    quark-anti-quark luminosities contributing to the 
     $W^+$ production in the proton-proton collisions at the c.m.s. energy of 
     $\sqrt{s}=14~{\rm TeV}$ (left panel) 
     and antiproton-proton collisions at the c.m.s. energy of 
     $\sqrt{s}=1.96~{\rm TeV}$ (right panel). 
       \label{fig:wzlum}
    }
  \end{center}
\end{figure}
%--------------------------------------------------------------------------------
In the present analysis we obtain
%-----------------------------------------------------------------------------------
\begin{eqnarray}
\langle xV(Q^2_0)\rangle  &=&  0.1646 \pm 0.0027~~~~{\rm (this~analysis)}\\
\langle xV(Q^2_0)\rangle  &=&  0.1610 \pm 0.0043~~~~{\rm N^3LO}~,
\label{BBGMOM}
\end{eqnarray}
%-----------------------------------------------------------------------------------
for $Q_0^2 = 4~{\rm GeV}^2$, where we combine in Eq.~(\ref{BBGMOM}) the value of the 
difference $x(u_v - d_v)$ obtained in Ref.~\cite{Blumlein:2006be} with the value for
%-----------------------------------------------------------------------------------
\begin{eqnarray}
\langle x[\bar{d} - \bar{u}]\rangle = 0.0072 \pm 0.0007
\end{eqnarray}
%-----------------------------------------------------------------------------------
found in the present analysis. In the above combination the correlation to the 
heavier 
flavor distributions is negligible.

The PDF-uncertainties given in 
Figures~\ref{fig:comp}-\ref{fig:comp2}
are defined by the uncertainties in the analyzed data and the 
uncertainties due to $m_c$, $m_b$, and the 
low-$x$ non-singlet exponent 
$a_\Delta$ as discussed in Section~\ref{sec4}.  
For the $c$- and $b$-quark distributions, the essential uncertainties are 
due to $m_c$ and $m_b$, respectively. At small $x$, however 
they are determined much more precisely than the strange sea distribution,  
which is widely unconstrained at $x\lesssim 0.01$ by the present data.
We now turn to some important inclusive processes at 
hadron colliders, for which we illustrate the impact of 
NNLO PDFs derived in the present analysis.
%---------------------------------------------------------------------------------
\subsection{\boldmath $W/Z$-boson Production}
\label{sec53}
%---------------------------------------------------------------------------------
\vspace{1mm}\noindent
The inclusive production cross sections of single $W^\pm$- and $Z$-bosons are
considered so-called standard candle processes at hadron colliders. 
The cross sections and distributions for these processes are calculated 
up to the NNLO~\cite{Hamberg:1990np,dyrap,Harlander:2002wh,Catani:2009sm} 
(see also Ref.~\cite{Blumlein:2005im} for the expressions in Mellin-space) 
which allows to reduce the theoretical uncertainty 
due to factorization and renormalization scale 
variation down to a few per cent.
With this theoretical accuracy provided the measurement of the $W^\pm/Z$-boson production rates 
can be used to monitor the luminosity of the collider. 
Moreover, a combination of the data on $W^\pm/Z$-production with the non-resonant 
Drell-Yan data allows to separate the quark distributions of different flavors 
with a very good accuracy, cf.~\cite{Dittmar:1997md}.
The quark-anti-quark luminosities contributing to $W^+$-production in $pp$ 
collisions are given by
%----------------------------------------------------------------------------------
\begin{eqnarray}
L_{q\bar{q}}=\tau\left[ q(\sqrt{\tau} e^Y,M_W)
\bar{q}(\sqrt{\tau} e^{-Y},M_W)
+\bar{q}(\sqrt{\tau} e^Y,M_W)
{q}(\sqrt{\tau} e^{-Y},M_W) \right],
\label{eqn:wzlum}
\end{eqnarray}
%----------------------------------------------------------------------------------
where $\tau=M^2_W/s$, $s$ denotes the c.m.s. collision energy 
squared, and 
$Y$ is the $W^+$ c.m.s rapidity. In Figure~\ref{fig:wzlum} we compare 
the luminosities of Eq.~(\ref{eqn:wzlum}) weighted by the corresponding 
Cabibbo-Kobayashi-Maskawa (CKM) matrix elements $V^2_{q_{i}\bar q_{j}}$
for different channels at the energy of the LHC with our NNLO 5-flavor 
PDFs used as input and  
\renewcommand{\arraystretch}{1.3}
\begin{table}[h]
\begin{center}
{\small
\begin{tabular}{|c|c|c|c|c|}   
\hline
\multicolumn{1}{|c|}{$\sqrt{s}$ [TeV]} &
\multicolumn{2}{c|}{this paper} &
\multicolumn{2}{c|}{MSTW \cite{Martin:2009iq}}  \\
\cline{2-5}
\multicolumn{1}{|c|}{ } &
\multicolumn{1}{c|}{$\sigma(\rm W^+ + W^-)$} &
\multicolumn{1}{c|}{$\sigma(Z)$} &  
\multicolumn{1}{c|}{$\sigma(\rm W^+ + W^-)$} &
\multicolumn{1}{c|}{$\sigma(Z)$} \\ 
\hline
$1.96~~(\bar p p)$ & $26.2 \pm 0.3$ & $7.73 \pm 0.08$ 
                   & $25.4 \pm 0.4$ & $7.45 \pm 0.13$
\\ \hline
$~~7~~~~(p p)$ & $98.8 \pm 1.5$ & $28.6 \pm 0.5$
              & & \\ \hline
$10~~~~(p p)$ & $145.6 \pm 2.4$ & $42.7 \pm 0.7$
              & $142.1 \pm 2.4$ & $42.5 \pm 0.7$
\\ \hline
$14~~~~(p p)$ & $207.4 \pm 3.7$ & $61.4 \pm 1.1$
              & $201.1 \pm 3.3$ & $61.0 \pm 1.0$
\\ \hline
\end{tabular}
}
\caption[]{\small The total $W^{\pm}$ and  $Z$-cross sections $[nb]$ at 
the Tevatron and LHC at the scale $\mu = M_{W/Z}$ (see Eq.~(\ref{eqn:wzpar}) for the other parameters)
with the PDFs and its estimated uncertainties from the 
present analysis and in comparison to results of Ref.~\cite{Martin:2009iq}.}
\label{tab:wzcs}
\end{center}
\normalsize
\end{table}
\renewcommand{\arraystretch}{1}
%----------------------------------------------------------------------------------
%
%\vspace*{-1cm}
\begin{equation}
V^2_{u\bar d}=V^2_{c\bar s}=0.9474,~~~~V^2_{u\bar s}=0.0509,~~~~~
M_W=80.398~{\rm GeV}~.
\label{eqn:wzpar}
\end{equation}
%----------------------------------------------------------------------------------
In the forward region of rapidity the main contribution comes from the $u$-$\bar{d}$ annihilation. 
In the central region the $c$-quark contribution is also essential. 
Therefore, the single $W^\pm$ cross section  measurement can be used to check the magnitude of the $c$-quark distribution. 
For the case of antiproton-proton collisions, the quark-anti-quark luminosities 
are similar to Eq.~(\ref{eqn:wzlum}); however, at Tevatron the valence 
$u$-quark contribution is dominating in the whole range of rapidity. 
%--------------------------------------------------------------------------------
%--------------------------------------------------------------------------------
The cross sections for $W^\pm/Z$-production at the scale $\mu = M_{W/Z}$ 
for the parameters in Eq.~(\ref{eqn:wzpar}), and $M_Z=91.188~{\rm GeV}$, 
and including the NNLO corrections of Refs.~\cite{Hamberg:1990np,Harlander:2002wh} are given in Table~\ref{tab:wzcs}. 
The quoted uncertainties are propagated from the uncertainties in the parameters of our 
PDFs, 
$\alpha_s$, $m_c$ and $m_b$, cf. Section~\ref{sec4}. 
They 
amount to $\sim 1\%$ at the Tevatron and  $\sim 2\%$ at the LHC. 
Comparing the present analysis to Ref.~\cite{Martin:2009iq} 
the results for Tevatron are at variance by $2\sigma_P$, while the same cross 
sections are obtained for $Z$-boson production at LHC energies.

%%%%%%%%%%%%%%%%%%%%%%%%%%%%%%%%%%%%%%%%%%%%%%%%%%%%%%%%%%%%%%%%%%%%%%%%%%%%%%%%%
\subsection{Top-quark Pair-Production}
\label{sec54}
%%%%%%%%%%%%%%%%%%%%%%%%%%%%%%%%%%%%%%%%%%%%%%%%%%%%%%%%%%%%%%%%%%%%%%%%%%%%%%%%%

\vspace{1mm}\noindent
The scattering cross section for hadroproduction of heavy quarks of mass $m_h$ 
is known exactly in QCD including radiative corrections at NLO~\cite{Nason:1987xz,Beenakker:1988bq,Bernreuther:2004jv,Czakon:2008ii}.
At NNLO approximate results based on the complete logarithmic dependence
on the heavy quark velocity $\beta = \sqrt{1-4m_h^2/\hat{s}}$ near threshold $\hat{s} \simeq 4m_h^2$ 
($\hat{s}$ being the partonic c.m.s. energy)
together with the exact dependence on the scale $\mu$ provide currently 
the best estimates~\cite{Moch:2008qy,Langenfeld:2009wd}.
\renewcommand{\arraystretch}{1.3}
\begin{table}[h]
\begin{center}
\begin{tabular}{|c|c|c|c|}   
\hline
$\sqrt{s}~({\rm TeV})$ &  this paper & MSTW2008  \\ \hline
$1.96~~(\bar p p)$ & $6.91\pm0.17$ 
                   & $7.04$ 
\\ \hline
$~~7~~~~(p p)$ & $131.3 \pm 7.5$ 
              & 160.5
\\ \hline
$10~~~~(p p)$ & $343\pm15$ 
              & $403$ 
\\ \hline
$14~~~~(p p)$ & $780\pm28$ 
              & $887$ 
\\ \hline
\end{tabular}
\caption[]{\small The total $t{\bar t}$-production cross sections $[pb]$ 
at the Tevatron and LHC for a pole mass of $m_t=173$~GeV at the scale $\mu=m_t$.
The results for the PDFs and its estimated uncertainties from the present analysis
are compared to the central values obtained using the 
PDFs of Ref.~\cite{Martin:2009iq}.}
\label{tab:ttbar}
\end{center}
\normalsize
\renewcommand{\arraystretch}{1.0}
\end{table}
%--------------------------------------------------------------------------------
At Tevatron, the cross section is most sensitive to the $q\bar q$-annihilation channel, 
with the luminosities $L_{ij}$ ordered in magnitude according to $L_{q \bar q} > L_{qg} > L_{gg}$.
At the LHC, on the other hand, the cross section receives the dominant
contribution from the $gg$-channel, in particular, from the gluon PDF 
in the region $x\approx 2.5 \cdot 10^{-2}$.
This makes the cross section for top-quark pair-production an interesting observable to investigate the gluon luminosity. 
Also the correlations of rates for $t\bar{t}$-pairs with other cross sections
can be studied quantitatively~\cite{Nadolsky:2008zw}.

Our cross sections for $t\bar{t}$-production are summarized in Table~\ref{tab:ttbar} 
for a pole mass of $m_t=173$ GeV.
We estimate the relative accuracy due to the PDF-fit for Tevatron by $\sim 3\%$, and for the LHC by $\sim 3.5-4.5\%$. 
With comparison of the cross sections obtained with the PDFs of Ref.~\cite{Martin:2009iq} 
we find agreement within $1\sigma_P$ for Tevatron.
For LHC energies, the results for the MSTW08 set are larger by $4\sigma_P$ due to 
a bigger value of $\alpha_s(M_Z^2)$ 
and the larger value of the gluon PDF in the partonic threshold region $\hat{s} \simeq  4 m_t^2$.
Note that the variation of the factorization and renormalization scale is not
considered here. It contributes separately to theoretical uncertainty (at NNLO
$\sim 3-4\%$ at Tevatron and LHC, see~\cite{Moch:2008qy,Langenfeld:2009wd} for details).

%%%%%%%%%%%%%%%%%%%%%%%%%%%%%%%%%%%%%%%%%%%%%%%%%%%%%%%%%%%%%%%%%%%%%%%%%%%%%%%%%
\subsection{Higgs Boson Production} 
\label{sec55}
%%%%%%%%%%%%%%%%%%%%%%%%%%%%%%%%%%%%%%%%%%%%%%%%%%%%%%%%%%%%%%%%%%%%%%%%%%%%%%%%%

\vspace{1mm}\noindent
Higgs-boson production is the most prominent signal at LHC and currently subject 
to intensive searches at Tevatron. 
The gluon-fusion channel (via a top-quark loop) is by far the largest production mode 
and known including 
the NNLO QCD corrections~\cite{Harlander:2002wh,Anastasiou:2002yz,Ravindran:2003um,Catani:2007vq}.

\begin{figure}[htb]
  \begin{center}
    \includegraphics[width=11.0cm]{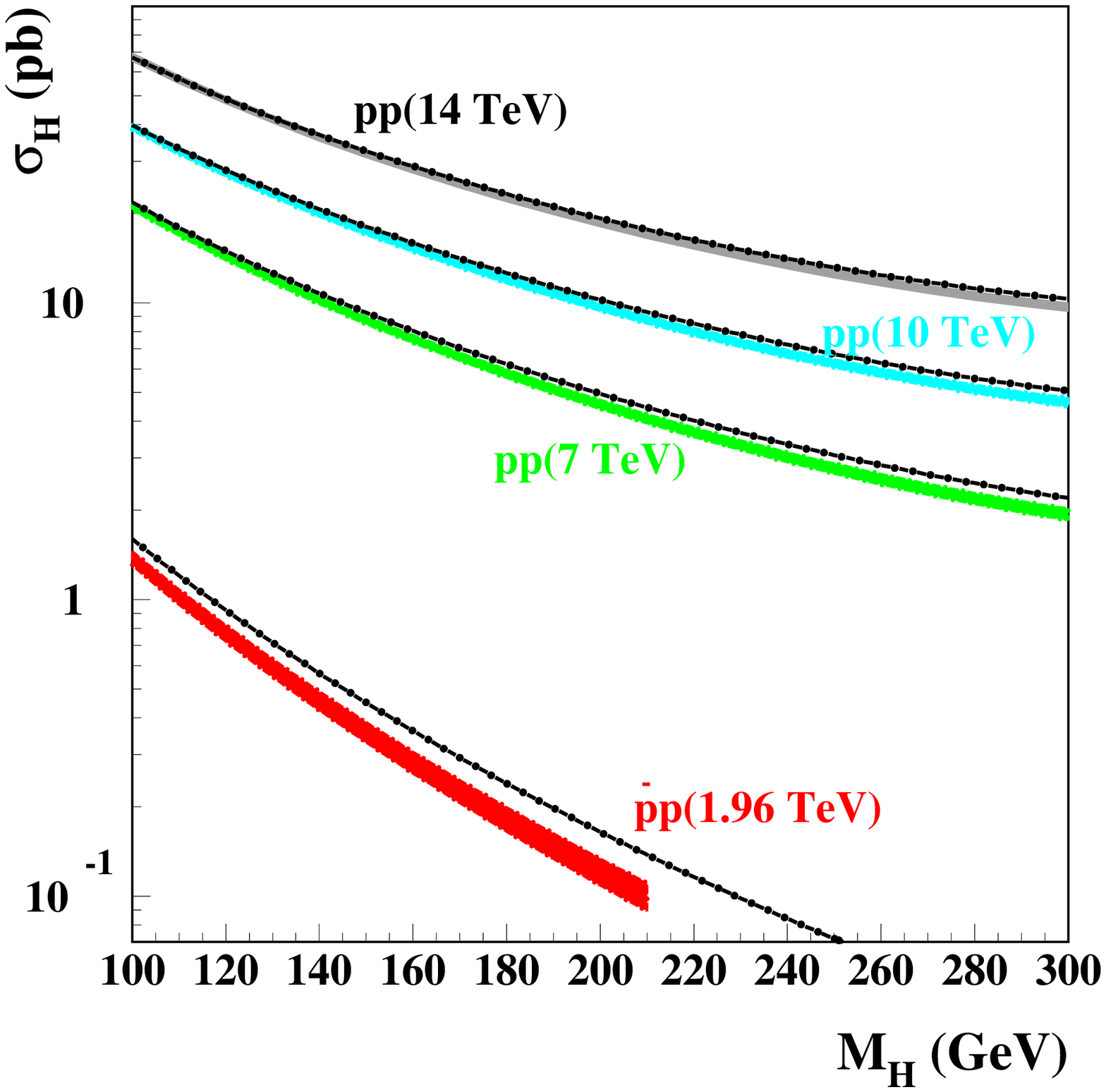}
    \vspace*{-5mm}
    \caption[]{ \small 
The $1\sigma_P$ error band for 
the Higgs-boson production cross sections $[pb]$ 
at Tevatron and the LHC at the scale of $\mu= M_H$  
employing the PDFs from the present analysis
(shaded area) in comparison with the central values for the case 
of PDFs of Ref.~\cite{Martin:2009iq} (dash-dotted lines).
      \label{fig:higgs}
    }
  \end{center}
\end{figure}

In Table~\ref{tab:higgs} the total production cross sections for the Higgs-boson 
are presented as a function of the Higgs-boson mass $m_H$ at Tevatron and for a series
of foreseen collision energies at the LHC (using $m_t=173$ GeV). 
The relative error from the PDF fit amounts to $5.5 - 10 \%$ 
at Tevatron and to $2.5 - 3 \%$ at the LHC at the higher energies and to $3.5-4.5\%$ 
at $\sqrt{s} = 7~{\rm TeV}$.
Again we do not consider the theoretical uncertainty due to the variation of the 
factorization and renormalization scale (typically amounting to $\sim 9 - 10\%$ at NNLO). 
In Figure~\ref{fig:higgs} we compare the production 
cross sections to the results obtained using the PDFs of Ref.~\cite{Martin:2009iq}.
The MSTW08 predictions yield higher values. For the LHC energies both analyses 
agree
at lower Higgs masses $M_H \sim 100$ GeV and a gradual deviation reaching $3
\sigma_P$ at $M_H = 300$ GeV of the MSTW08 values is observed. Our values at
Tevatron are lower than those of MSTW08 by $\sim 3 \sigma_P$ in the whole mass
range. 
\renewcommand{\arraystretch}{1.3}
\begin{table}[h]
\begin{center}
{\footnotesize
\begin{tabular}{|c|c|c|c|c|}   
\hline
\multicolumn{1}{|c|}{$m_H/{\rm GeV}$  } &
\multicolumn{1}{c|}{Tevatron} &
\multicolumn{1}{c|}{LHC 7 TeV}  &
\multicolumn{1}{c|}{LHC 10 TeV}  &
\multicolumn{1}{c|}{LHC 14 TeV}  \\
\hline
\hline
  100.  & $1.381 \pm 0.075$ & $21.19  \pm 0.58$ & $39.17  \pm  1.05$ & $67.28  \pm 1.77$  \\
  110.  & $1.022 \pm 0.061$ & $17.30  \pm 0.49$ & $32.52  \pm  0.88$ & $56.59  \pm 1.51$  \\
  120.  & $0.770 \pm 0.049$ & $14.34  \pm 0.41$ & $27.38  \pm  0.72$ & $48.25  \pm 1.23$  \\
  130.  & $0.589 \pm 0.041$ & $12.03  \pm 0.36$ & $23.33  \pm  0.61$ & $41.60  \pm 1.07$  \\
  140.  & $0.456 \pm 0.033$ & $10.21  \pm 0.31$ & $20.08  \pm  0.55$ & $36.23  \pm 0.92$  \\
  150.  & $0.358 \pm 0.028$ & $8.75  \pm 0.27$ & $17.45  \pm  0.48$ & $31.83  \pm 0.82$  \\
  160.  & $0.283 \pm 0.024$ & $7.56  \pm 0.24$ & $15.29  \pm  0.43$ & $28.20  \pm 0.72$  \\
  170.  & $0.226 \pm 0.020$ & $6.59  \pm 0.21$ & $13.51  \pm  0.37$ & $25.16  \pm 0.65$  \\
  180.  & $0.183 \pm 0.017$ & $5.78  \pm 0.19$ & $12.01  \pm  0.35$ & $22.60  \pm 0.60$  \\
  190.  & $0.148 \pm 0.014$ & $5.11  \pm 0.17$ & $10.75  \pm  0.31$ & $20.44  \pm 0.53$  \\
  200.  & $0.121 \pm 0.013$ & $4.55  \pm 0.16$ & $9.69  \pm  0.28$ & $18.59  \pm 0.49$  \\ 
  210.  &                   & $4.07  \pm 0.15$ & $8.78  \pm  0.26$ & $17.01  \pm 0.44$  \\
  220.  &                   & $3.67  \pm 0.14$ & $8.00  \pm  0.24$ & $15.64  \pm 0.42$  \\
  230.  &                   & $3.32  \pm 0.13$ & $7.33  \pm  0.22$ & $14.46  \pm 0.38$  \\
  240.  &                   & $3.02  \pm 0.12$ & $6.75  \pm  0.21$ & $13.44  \pm 0.37$  \\
  250.  &                   & $2.77  \pm 0.11$ & $6.25  \pm  0.20$ & $12.55  \pm 0.35$  \\
  260.  &                   & $2.55  \pm 0.10$ & $5.82  \pm  0.19$ & $11.79  \pm 0.32$  \\
  270.  &                   & $2.36  \pm 0.10$ & $5.45  \pm  0.18$ & $11.12  \pm 0.31$  \\ 
  280.  &                   & $2.19  \pm 0.10$ & $5.13  \pm  0.17$ & $10.56  \pm 0.30$  \\
  290.  &                   & $2.06  \pm 0.09$ & $4.86  \pm  0.17$ & $10.08  \pm 0.29$  \\
  300.  &                   & $1.94  \pm 0.09$ & $4.63  \pm  0.16$ & $ 9.69  \pm 0.28$  \\
\hline
\end{tabular}
}
\caption[]{\small The total cross sections for Higgs-boson production $[pb]$ at 
Tevatron and the LHC at the scale $\mu = M_H$ with the uncertainties estimated from the fit results in 
the present analysis.}
\label{tab:higgs}
\end{center}
\normalsize
\renewcommand{\arraystretch}{1.0}
\end{table}
%--------------------------------------------------------------------------------
At the LHC energies the difference can be attributed to different 
gluon PDFs and values for $\alpha_s$.
The cross sections take very similar values for light Higgs masses, but beyond 
scales $\mu^2 \sim 10^4$ GeV$^2$ the values obtained with MSTW08 are larger.
%--------------------------------------------------------------------------------
%--------------------------------------------------------------------------------

%%%%%%%%%%%%%%%%%%%%%%%%%%%%%%%%%%%%%%%%%%%%%%%%%%%%%%%%%%%%%%%%%%%%%%%%%%%%%%%%%
\section{Conclusions} 
\label{sec6}
%%%%%%%%%%%%%%%%%%%%%%%%%%%%%%%%%%%%%%%%%%%%%%%%%%%%%%%%%%%%%%%%%%%%%%%%%%%%%%%%%

\vspace*{1mm}\noindent
The precision of the DIS world data has reached a level which requires
NNLO analyses to determine the PDFs and to measure the 
strong coupling constant $\alpha_s(M_Z^2)$. This also applies to the most
prominent scattering processes at hadron colliders such as the Drell-Yan process, 
$W^\pm$-, $Z$-boson, Higgs-boson and top-quark pair-production.
In the present analysis we have performed an NNLO fit to the DIS world data, 
Drell-Yan- and di-muon data along with a careful study of the heavy flavor effects 
in the DIS structure function $F_2$. 
In the analysis we have taken into account correlated errors whenever available. 
In total, 25 parameters have been fitted yielding a positive semi-definite covariance matrix. 
With this information one may predict the error with respect to the PDFs, 
$\alpha_s(M_Z^2)$, $m_b$ and $m_c$ for hard cross sections measured, including all 
correlations. 
For applications to hadron collider processes we have determined 3-, 4- and
5-flavor PDFs within the GMVFN scheme applying the BMSN description.
We have performed a detailed study of the heavy flavor contributions to deep inelastic 
scattering comparing to experimental data. We have compared to different treatments used in 
the literature and found that both the FFN scheme and the BMSN scheme 
yield a concise description 
of the DIS data at least for the kinematic range of HERA, and that no modifications of 
these renormalization group-invariant prescriptions are needed. 
In the present analysis we have obtained $\alpha_s(M_Z^2)$ 
with an accuracy of $\approx 1.5\%$. 
The values quoted in Eqs.~(\ref{eq:alphas-FFN}) and (\ref{eq:alphas-VFN}) are
found to be in very good agreement with the non-singlet analysis of Ref.~\cite{Blumlein:2006be}, 
which relied on a sub-set of the present data only, and with the results of Ref.~\cite{JR}.
The central value of $\alpha_s(M_Z^2)$ steadily converges going from LO to NLO to  
NNLO, or even to N$^3$LO in the non-singlet case \cite{Blumlein:2006be}. The 
differences in the central values (determined at $\mu^2 = Q^2$) provide a good estimate of the
remaining theory errors. 
It is very hard to achieve a better accuracy on $\alpha_s(M_Z^2)$ than obtained at the moment, 
given the theoretical uncertainties (reaching values around $\sim 0.7 \%$),
which arise from the difference between the FFN and 
BMSN scheme, from quark mass effects, from 4-loop effects in the strong coupling constant 
from (the yet unknown) effect of the 4-loop singlet anomalous dimensions, or from remainder higher 
twist effects and so on.
However, potential 
high-luminosity measurements planned at future facilities like EIC \cite{EIC}, 
requiring an excellent control in the {\it systematics}, may provide future 
challenges to the precision on the theoretical side. 

We have discussed the NNLO PDFs of the present fit and compared to other global analyses.
A comparison to the results of MSTW08 in the region $\mu^2 = 4$ to $10^4$ GeV$^2$ 
show that smaller values for the light PDFs for lower values of $x$ are obtained in Ref.~\cite{Martin:2009iq}. 
Moreover, the gluon distribution of Ref.~\cite{Martin:2009iq} at low scales $\mu^2 = 4$ GeV$^2$ 
does strongly deviate from ours turning to negative values at $x \sim 5 \cdot 10^{-5}$. 
At large values of $x$ the gluon distribution of Ref.~\cite{Martin:2009iq} is slightly larger than ours.
Somewhat smaller values are also obtained for the $c$- and $b$-quark distributions.
The PDFs obtained in Ref.~\cite{JR} agree very well with the results of the present 
analysis.

We have illustrated the implications of the PDFs for standard candle processes,
such as $W^\pm$- and $Z$-boson production at hadron colliders.
Comparison to MSTW08 yields a $2\sigma_P$ lower result for Tevatron and better agreement 
is obtained for the LHC energies. 
Conversely, the inclusive $t\bar{t}$ production cross section of both analyses agree 
at Tevatron energies, but for the LHC larger results by $\sim 2\sigma_P$ are obtained with MSTW08. 
For the inclusive Higgs-boson production cross section at Tevatron the PDFs
of MSTW08 yield a $3\sigma_P$ larger value in the whole mass range,
while for LHC energies both predictions agree for masses $M_H \sim 100$ GeV, 
and MSTW08 gives by $3\sigma_P$ larger values for $M_H \sim 300$ GeV.
Of course, all observed differences have to be considered in view of the statistical and 
systematic accuracies finally to be obtained in the experimental measurements.    

The PDFs of the present analysis allow for detailed simulations of the 
different inclusive processes at the LHC and are of central importance in monitoring the luminosity.
Precision measurements of inclusive processes at hadron colliders open up the 
opportunity to further refine the understanding of the PDFs of nucleons. 
This applies to both, the final analyses at Tevatron and the future measurements at the LHC. 
During the last years our understanding of PDFs has steadily improved at the NNLO level 
and upcoming high luminosity data from hadron colliders will continue in this direction.

Grids, which allow fast access to our 3-, 4-, and 5-flavor PDFs 
in a wide range of $x$ and $Q^2$ 
(including the PDF uncertainties considered)
are available online at~\cite{www}.

\noindent
{\bf Acknowledgments.}\\
We would like to thank P. Nadolsky, D. Renner, E. Reya, J. Smith, 
W.J. Stirling and R. Thorne for discussions. This work was supported in 
part by DFG
Sonderforschungsbereich Transregio 9, Computergest\"utzte Theoretische
Teilchenphysik, the RFBR grant 08-02-91024, Studienstiftung des Deutschen 
Volkes, the European Commission MRTN HEPTOOLS under Contract No. 
MRTN-CT-2006-035505. 

\newpage
%----------------------------------------------------------------------------------------

%----------------------------------------------------------------------------------------
%%%%%%%%%%%%%%%%%%%---------------------------------------------------------------
\end{document}